\documentclass[twocolumn,prd,nofootinbib,aps,prl,floats,floatfix,amsmath,amssymb,longbibliography,secnumarab
ic]{revtex4-1} %

\usepackage[final]{graphicx}
\usepackage{hyperref}
\usepackage{amsmath}
\usepackage{bbm}
\usepackage{amsfonts}
\usepackage{amssymb}
\usepackage{latexsym}
\usepackage{graphicx}
\usepackage[english]{babel}
\usepackage{multirow}
\usepackage{float}
\usepackage{url}
\usepackage{slashed}
\usepackage{xcolor} 
\usepackage[utf8]{inputenc}
\usepackage{verbatim}
\usepackage{stackengine}
\usepackage{cancel}
\usepackage{accents}
\usepackage{array}
\usepackage{physics}
\usepackage{bm}

\def\Abs#1{\big|#1\big|}
\newcommand{\lp}{\left(}
\newcommand{\rp}{\right)}
\newcommand{\lbr}{\left[}
\newcommand{\rbr}{\right]}
\newcommand{\lb}{\left\lbrace}
\newcommand{\rb}{\right\rbrace}
\newcommand{\lc}{\left[}
\newcommand{\rc}{\right]}
\newcommand{\be}{\begin{equation}}
\newcommand{\ee}{\end{equation}}
\newcommand{\ba}{\begin{array}}
\newcommand{\ea}{\end{array}}
\newcommand{\bea}{\begin{eqnarray}}
\newcommand{\eea}{\end{eqnarray}}

\newcommand{\nn}{\nonumber}

\newcommand{\ab}{{\rm (abs)}}
\newcommand{\sn}{{\widetilde{N}}}
\newcommand{\slp}{{\widetilde{L}}}
\newcommand{\ho}{{\widetilde{H}}}

\newcommand{\gw}{{\rm gw}}
\newcommand{\SNR}{{\rm SNR}}
\usepackage{slashed}

\renewcommand{\c}{{\rm c}}
\newcommand{\s}{{\rm s}}
\newcommand{\ch}{{\rm ch}}
\newcommand{\sh}{{\rm sh}}

\begin{document}

\title{PeV-scale leptogenesis, gravity waves and black holes\\ from a SUSY-breaking phase transition}
\author{James M.\ Cline}
\author{Benoit Laurent}
\author{Jean-Samuel Roux}
\affiliation{McGill University, Department of Physics, 3600 University St.,
Montr\'eal, QC H3A2T8 Canada}
\author{Stuart Raby}
\affiliation{The Ohio State University, 191 W. Woodruff Ave., Columbus, OH 43210}

\begin{abstract}
Supersymmetry is a highly motivated theoretical framework, whose scale of breaking may be at PeV energies, to explain null searches at the Large Hadron Collider.  SUSY breaking through a first order phase transition may have occurred in the early universe, leading to potential gravitational wave signals.  Constructing a realistic model for gauge-mediated
supersymmetry breaking, 
we show that such a transition can also induce masses for heavy
right-handed neutrinos and sneutrinos, whose CP-violating decays give  leptogenesis at the PeV scale, and a novel mechanism of neutrino mass
generation at one loop.  For the same models we predict the possible gravity wave signals, and we study the possibility of production of
primordial black holes during the phase transition.
\end{abstract}
\maketitle

\section{Introduction}

First order phase transitions (FOPTs) in the early universe have taken on renewed interest, since the sound waves produced during the transition may produce primordial gravitational waves (GWs) that could be observed by upcoming experiments, including LISA \cite{Audley:2017drz,Caprini:2015zlo,Caprini:2019egz}, DECIGO \cite{Seto:2001qf,Sato:2017dkf} and BBO \cite{Crowder:2005nr,Harry:2006fi}.  Moreover
FOPTs can produce nonequilibrium effects such as are needed for baryogenesis, notably when the electroweak phase transition is involved.  Possible correlations between the gravity wave signals and successful electroweak baryogenesis have been widely studied
\cite{Dorsch:2016nrg,Vaskonen:2016yiu,Cline:2021iff}.  However it is also possible to have analogous scenarios for other kinds of phase transitions, notably where lepton number is spontaneously broken 
\cite{Long:2017rdo,Huang:2022vkf,Dasgupta:2022isg}.

It is interesting to consider supersymmetric (SUSY) generalizations of this framework.
Although there is not yet experimental evidence for SUSY, it figures prominently in string-theoretic completions of the standard model, which may be the most promising means of reconciling quantum mechanics with gravity.  Current constraints from ATLAS and CMS suggest that SUSY is broken well above the TeV scale \cite{Adam:2021rrw}, disappointing hopes that naturalness of the weak scale would imply a low scale of SUSY breaking.  Nevertheless, it has been argued that SUSY breaking even at the PeV scale need not be severely fine tuned \cite{Wells:2004di}.  Moreover, it has been shown that hidden-sector SUSY breaking at the PeV scale, through a FOPT, can readily produce
observable gravity waves, with novel features in the high-temperature dynamics of the phase transition \cite{Craig:2020jfv}.

In the present work, we propose an explicit model of gauge-mediated
SUSY breaking that can yield successful leptogenesis, through the out-of-equilibrium decays of heavy right-handed neutrinos and their superpartners.  
Usually the scale of leptogenesis is decoupled from that of SUSY breaking, but in this work we consider
a model where the two are tied together; we assume that the same field that breaks SUSY also produces a large lepton-violating mass
for heavy right-handed neutrinos.
We take advantage of the near-degeneracy of heavy right-handed neutrinos in a minimal flavor violation (MFV) hypothesis to resonantly enhance the CP asymmetry of their decays, a regime known as resonant leptogenesis \cite{Pilaftsis:2003gt}, and we extend it to the decays of sneutrinos. The resonance allows the model to yield the observed baryon asymmetry despite strong washout effects due to inverse decays. 

We found that O'Refeartaigh-like models of SUSY breaking require at least two right-handed neutrino superfields per lepton flavor  to have a nonvanishing vacuum energy in this scenario, in contrast to the usual assumption of just a single family. The level repulsion between these two fields pushes the scale of leptogenesis down to $\mathcal{O}(10^5-10^6 {\rm\ GeV})$, while generating light neutrino masses from one-loop effects at the $\mathcal{O}(10^8 {\rm\ GeV})$ scale.

We start with a description of the general framework of
gauge-mediated hidden-sector SUSY breaking in Section \ref{sec:framework}, and there introduce a specific model.
In Section \ref{sec:FOPT} we construct the finite-temperature effective potential needed for analyzing the first-order SUSY-breaking phase transition that produces gravitational waves (GWs).  Section \ref{sec:lepto} describes the generation of the lepton asymmetry that gives rise to baryogenesis, via resonant leptogenesis, as well as the one-loop mechanism for neutrino mass generation.  In Section \ref{sec:signatures} we describe two additional cosmological consequences: the production of potentially observable gravity waves, and constraints arising from production of primordial black holes at the end of the 
phase transition.  Conclusions are given in Section \ref{sec:conclusion}.
Details of the one-loop effective potential including finite temperature are given in the appendix.

\section{Framework}\label{sec:framework}
Ref.\ \cite{Craig:2020jfv} considered scenarios of gauge-mediated
SUSY breaking from a hidden sector, in which a chiral pseudomodulus superfield $X$ gets a vacuum expectation value (VEV) in its scalar component $x$, through a first order phase transition.  The desired shape of the potential $V(x)$, with a barrier separating the true and false vacua, is generated by four vectorlike messenger superfields, $\Phi$, $\Phi'$, $\overline\Phi$ and $\overline\Phi'$,
which also communicate SUSY breaking to the visible sector, as in the O’Raifeartaigh model \cite{ORaifeartaigh:1975nky}.  They are SM singlets, but they carry charges $+1$, $+1$, $-1$, $-1$ respectively, under a U(1)$_D$ gauge symmetry, which allows for a Fayet-Iliopoulos (FI) $D$-term that can enhance the gravity wave signal.

This minimal particle content is sufficient to produce a FOPT, but to realistically embed the standard model (in the guise of the
minimal supersymmetric standard model, MSSM), one can unify the SM gauge group in SU(5), and include vectorlike messengers analogous to the $\Phi$ fields but in the fundamental representation of SU(5), which we denote by $5_M$, $5'_M$, $\overline 5_M$, $\overline 5'_M$.  In addition, we introduce right-handed neutrino superfields $N_i$, $N_i'$ whose coupling to $X$ leads to lepton number violating interactions after SUSY breaking, enabling leptogenesis.  We note that it is not strictly necessary to 
unify the SM gauge group to SU(5), but the  $5_M$
mediators provide a simple  way of communicating SUSY breaking to the SM.  We can then consider SU(5) as an approximate global symmetry, broken by the gauge couplings of the SM. It enlarges to SU(6) by ignoring the U(1)$_D$ coupling and 
appropriately combining the $\Phi$ and $5_M$ fields:
\bea
    \Psi &=& (\Phi,\, 5_M)^{T},\quad \Psi' = (\Phi',\, 5'_M)\,,\nn\\
    \overline\Psi &=& (\overline{\Phi},\, \overline 5_M)^{T},\quad \overline\Psi' = (\overline{\Phi}',\, \overline 5'_M)\,.
\eea
The various fields and their transformation properties are summarized in Table
\ref{tab1}.

\begin{table}
\centering
\begin{tabular}{ | c | c | c | c | c | c|c|c|} 
 \hline
 Field & $R$ & U(1)$_D$ &  SU(5) & SU(2)$_L$ & U(1)$_y$ & $L$ & $\mathbb{Z}^L_2$\\
 \hline
 $X$ & $+2$ & 0 & 1 & 1 & 0&0&$+1$\\
 \hline
 $\Phi$ & 0 &  $+1$ & 1 & 1 & 0&0&$+1$\\
 $\overline{\Phi}$ & $+2$ & $-1$ & 1 & 1 & 0&0&$+1$\\
$ \Phi'$ & $+2$ & $+1$ &  1 & 1 & 0&0&$+1$\\
$\overline{\Phi}'$ & $0$ & $-1$ & 1 & 1 & 0&0&$+1$\\
\hline
$5_M$ & $0$ &  $0$ & 5 & * & $+y$&0&$+1$\\
$\overline{5}_M$ & $+2$ &  $0$ & $\overline 5$ & * & $-y$&0&$+1$\\
$5'_M$ & $+2$ &  $0$ & 5 & * & $+y$&0&$+1$\\
$\overline{5}'_M$ & $0$ &  $0$ & $\overline 5$ & * & $-y$ &0&$+1$\\
\hline
$N_i$ & $0$ & 0 & 1 & 1 & 0 &$-1$&$-1$\\
$N'_i$ & $+2$ & 0 & 1 & 1 & 0&$+1$&$-1$\\
\hline
$L_\alpha$ & $2$ & 0 & 1 & 2 & $-1$&$+1$&$-1$\\
$H_u$ & $0$ & 0 & 1 & 2 & $+1$&0&$+1$\\
\hline
\end{tabular}
\caption{Superfield content and their charges.  $*$ denotes that each $5$-plet contains an SU(2)$_L$ doublet (as well as an SU(3)$_c$ triplet).   The hypercharge $\pm y$ of the $5_M$ mediators is undetermined.}
\label{tab1}
\end{table}

We hypothesize that heavy sterile neutrinos get a signficant
contribution to their masses from spontaneous SUSY breaking, to 
provide a link between the phase transition and leptogenesis.
In our model, this comes from a superpotential term $X N_i N_j$.   However, as discussed below, this by itself would lead to the sneutrino fields $\tilde N_i$ getting VEVs and yielding a SUSY-preserving vacuum.  Such an undesired outcome can be avoided by the inclusion of two types of heavy neutrino fields $N_i$ and $N'_i$, carrying opposite lepton number,
allowing us to generate light neutrino masses at one loop, by $Z$ and $\tilde Z$ exchange.\footnote{Qualitatively similar one-loop mechanisms were considered in Refs.\ \cite{Ma:2006uv,Mohapatra:2022tgb}.} 

In addition to the fields shown, the MSSM contains the quark doublets $Q_\alpha$, the down-type Higgs $H_d$, and the SU(2) singlet quarks and leptons,
$U_\alpha$, $D_\alpha$, $E_\alpha$.  Although they do not play an immediate role in the present study, for completeness we suggest a possible set of $R$-symmetry charges for them, that would be compatible with SU(5) gauge coupling unification. 
The fields $Q_\alpha$, $U_\alpha$, $E_\alpha$  in the 10 dimensional representation have $R(10)=1$, while $D_\alpha$ and $L_\alpha$ in the $\bar 5$ have $R(\bar 5)=2$,  and $R(H_d)=-1$.  The $\mu$ term  ($\mu H_u H_d$) would be nonrenormalizable, requiring additional fields with total $R = 3$.  

The resulting superpotential, which can realize both spontaneous SUSY breaking and leptogenesis, is
\bea
W &=& W_{MSSM} - FX + \lambda X \Psi\overline\Psi'
 + m(\Psi\overline\Psi + \Psi'\overline\Psi')\nn\\
&+& \frac{\lambda'_{ij}}{2} X N_i N_j + M_{ij} N_i N'_j + Y_{i\alpha} \epsilon_{a b} N_i L_\alpha^a H_u^b\,, \label{superpot}
\eea 
where $W_{MSSM}$ is the superpotential of the MSSM,  $H_u$ is the up-type Higgs doublet, $\alpha$ is the flavor index for the SM fields, and $i$ is that for the right-handed neutrino superfields.  At least two flavors of $N_i$ are needed for CP-violation leading to leptogenesis and to explain light neutrino oscillations; for simplicity we will assume only two such flavors.  Terms of the form $N'NN$ and $F' N'$ are forbidden by a discrete symmetry $\mathbb{Z}_2^L$ under which all fields acquire a sign $(-1)^L$, where $L$ is the lepton number. This ``leptonic parity'' could be considered as a remnant of the broken $U(1)_L$ global symmetry. 

To construct a predictive and economical model, we will assume a version of MFV, {\it i.e.,} that SO($2$) symmetry of the  sterile
neutrino flavors  (labeled by $i$ for both $N_i$ and $N'_i$)
in the superpotential
is broken only by the Yukawa couplings $Y_{i\alpha}$,
so that at leading order  $\lambda'_{ij} = \lambda'\delta_{ij}$ and $M_{ij} = M\delta_{ij} + i M'\epsilon_{ij}$, using the invariant tensors of SO(2).
At one loop, corrections are generated involving $\delta Y^2_{ij}\equiv
Y_{i\alpha}Y^\dagger_{\alpha j}/(16\pi^2)$ \cite{Martin:1993zk}:
\bea
    \lambda'_{ij} &=& \lambda' \left[\delta_{ij} + c_1\,\delta Y^2_{ij}\right]\,,\nn\\ 
    M_{ij} &=& \left(\delta_{ik}+ c_2\,\delta Y^2_{ik}\right)
    \left[M\delta_{kj} + iM'\epsilon_{kj}\right]\,,
\eea
where $c_{1,2}$ are constants of order unity.  Ignoring the small corrections of order $Y^2/16\pi^2$, the eigenvalues of $M_{ij}$ are $M\pm M'$.

As we will show, the model requires $\delta Y^2_{ij}\sim10^{-8}$ (that is, $Y^2\sim 10^{-6}$) to radiatively generate neutrino mass of order $\sim0.05$ eV. Therefore, Yukawa corrections to the mass eigenvalues are negligible for $M'\gtrsim 10^{-7}M$ only.  It is technically natural to assume that $M'\ll M$ so that the two states, with masses $M\pm M'$, are nearly degenerate.   It will be shown in Section \ref{sec:lepto} that this leads to the quasi-resonant enhancement of the $CP$ asymmetry of neutrino and sneutrino decays. In this work, we will fix $M'=10^{-7}M$ to maximize the effect of this resonance. On the other hand, the Yukawa interactions contain CP-violating phases, that can lead to the CP asymmetries needed for leptogenesis, as we will show.

SUSY is spontaneously broken when the scalar component of $X$ gets its vacuum expectation value (VEV). The scale
\be
    \mu = 
    \lambda'\langle\tilde X\rangle \equiv {\lambda'\over 
    \sqrt{2}} \langle x\rangle
    \label{mudef}
\ee
will play an important role in the following.  It determines the
mass of the heavy sterile neutrinos $N_i$, in the regime $M\ll\mu$ where their mixing with $N'_i$ is suppressed.  Then $N'_i$ is relatively light, with mass $\sim M^2/\mu$.  We will see that this is the favorable regime for obtaining successful leptogenesis.  $\langle x\rangle$ is determined 
by the microscopic model parameters as
estimated in Eq.\ (\ref{xestimate}) below.

Another frequently occurring scale is the effective $F$ term
\be
     F_X\equiv F-\lambda \langle\widetilde{\Phi} \widetilde{\overline{\Phi}}'\rangle
\ee
at the minimum of the potential, where the fields $\widetilde{\Phi}$,
$\widetilde{\overline\Phi}'$ obtain VEVs, as described below.   It quantifies the
spontaneous breaking of SUSY, and appears in the gluino mass arising from gauge-mediated SUSY breaking,
\be \label{eq:mgluino}
m_{\tilde g} = N_m  \frac{g_s^2}{(4\pi)^2} \frac{F_X}{\langle\widetilde X\rangle}s_M\,C_{RG}\,, 
\ee 
where $g_s$ is the strong coupling at the scale $\langle\widetilde X\rangle$, and $N_m=2$ is the number of pairs of messenger $SU(3)$ triplets. The factor $s_M\equiv\frac16[\lambda F/m^2]^2$ expresses the effect of gaugino screening \cite{Cheung:2007es,Komargodski:2009jf,Cohen:2011aa}.\footnote{The screening suppression comes from the first subleading term in Eq.\ (7) of Ref.\ \cite{Martin:1996zb}, where the leading term is
subtracted by the gaugino mass counterterm.}  The correction $C_{RG}$ accounts for the renormalization group (RG) relation of the soft mass to the pole mass, from Eq.\ (8.3.3) of Ref.\ \cite{Martin:1997ns}.
The current
experimental limit of $m_{\tilde g} > 2.3\,$TeV \cite{ATLAS:2020syg} will provide a significant constraint on the
parameter space favorable for gravity wave production, in the following. Although we only consider the minimal model with $N_m=2$, one could always consider a model with more sets of messenger fields, which would increase gaugino masses with little impact on the FOPT dynamics or leptogenesis.

\subsection{Scalar potential}

The $F$-term potential arising from the superpotential (\ref{superpot}) includes the terms
\bea
  \label{VFeq}
    V_F &\supset& \abs{F - \frac{\lambda'}{2} \sn_i \sn_i
-\lambda \widetilde\Psi\widetilde{\overline\Psi}'}^2 
+ \left|m\right|^2\left(|{\widetilde\Psi}|^2+|{\widetilde{\overline\Psi}'}|^2\right)\nn\\
    &+&  \abs{\lambda \widetilde X \widetilde{\Psi} + m\widetilde{\Psi}'}^2 + \abs{\lambda \widetilde X \widetilde{\overline\Psi}' + m\widetilde{\overline\Psi}}^2\\
    &+& \abs{\lambda'\widetilde X\sn_i+ M_i \sn'_i
     + \epsilon_{a b} Y_{i\alpha} \widetilde L_\alpha^a H_u^b}^2
    + \abs{M_{i} \sn_i  }^2\nn
\eea
where tilde denotes the scalar component of the superfield (except later for $\widetilde H_u$, where it is the higgsino).

In addition, there is a $D$-term potential from the U(1)$_D$ interaction, including a Fayet-Iliopoulos term,
\be
    V_D = {g^2\over 2}\left({D\over g} + |\widetilde\Phi|^2 + 
    |\widetilde\Phi'|^2 - |\widetilde{\overline\Phi}|^2  -|\widetilde{\overline{\Phi}}'|^2
        \right)^2
\ee
where $g$ is the U(1)$_D$ gauge coupling.
We do not write the $D$ terms of the $5_M$ fields associated with their
SM charges, which only ensure that these fields have vanishing VEVs 
at the minimum of the potential.  Therefore the $\Psi$ fields in 
Eq.\ (\ref{VFeq}) can be replaced by their corresponding $\Phi$ components, for the purposes of minimizing the potential.

At zero temperature, there is a metastable minimum of $V=V_F+V_D$ at $\sn = \sn' = \widetilde\Psi= \widetilde\Psi'=0$, provided the conditions $M_i^2 > \lambda' F$  and $m^4 > \lambda^2 F^2 + g^2 D^2$ are satisfied. At tree level, $\widetilde X$ is a flat direction near this local minimum. However, one-loop corrections to the scalar potential slightly lift the flat direction, making $\widetilde X=0$ the local minimum.

If $D>F$, the true minimum of the tree-level potential is a runaway solution, $\widetilde X\to \infty$ and $\widetilde{\Phi},\widetilde{\overline{\Phi}}'\to0$, with $\widetilde{\Phi}' = -(\lambda/m) \widetilde X \widetilde{\Phi}$, $\widetilde{\overline\Phi} = -(\lambda/m) \widetilde X \widetilde{\overline \Phi}'$ taking finite values that satisfy
\be\label{eq:runaway}
\big|\widetilde{\Phi}'\big|^2 - \big|\widetilde{\overline\Phi}\big|^2 +\frac{D}{g} = 0\,.
\ee 
Therefore at the minimum of $V$, $V_D$ vanishes while $V_F=\abs{F}^2$. However, as we will show, loop corrections shift the minimum to finite values of $\widetilde X,\ \widetilde{\Phi}$ and $\widetilde{\overline{\Phi}}'$.

\begin{figure}[t]
    \centering
    \includegraphics[width=1\linewidth]{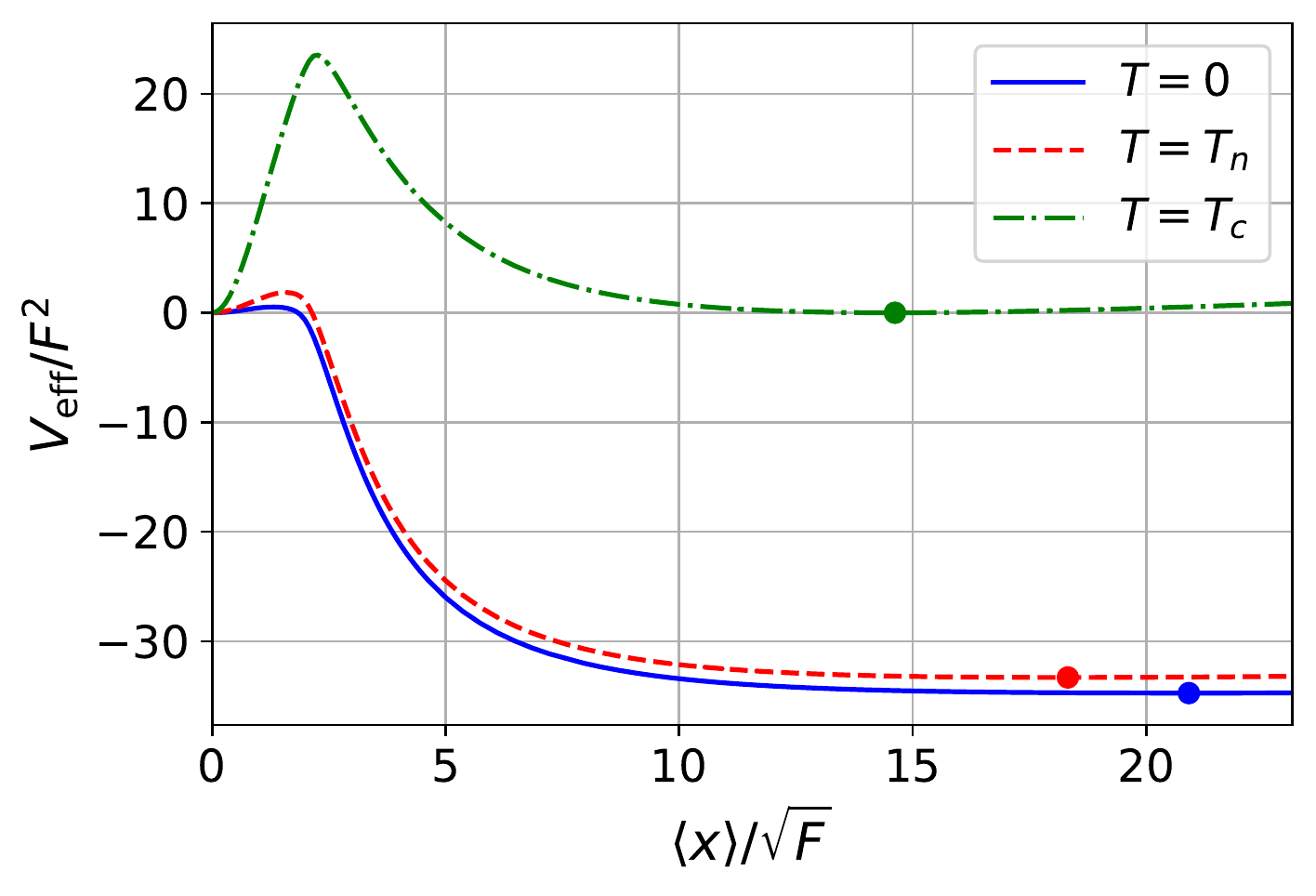}
    \caption{Typical profiles of $V_{\rm eff}(x)$ at $T=0,\, T_n\ \mathrm{and}\ T_c$. The points show the position of the true vacuum. Model parameters are fixed at $\lambda=4$, $D=8.6F$, $g=0.1$ and $m=\sqrt{5.3F}$.}
    \label{fig:Veff}
\end{figure}

At one loop, $V$ receives a Coleman-Weinberg correction through the field-dependent masses $m^2_i(x,\phi_1,\phi_2)$, where for brevity we define $x = \sqrt{2}\Abs{\widetilde X}$, $\phi_1 = \sqrt{2}\Abs{\widetilde{\overline\Phi}}$
and $\phi_2 = \sqrt{2}\Abs{\widetilde{\Phi}'}$. Expressions for the masses and $V_{CW}$ are given in Appendix \ref{appA}.  We define an effective
potential for the modulus alone, $V_{\rm eff}(x)$, by setting $\phi_1$ and $\phi_2$ to the values that minimize $V$ for fixed $x$.  Typical behaviors for $V_{\rm eff}(x)$ at different temperatures are illustrated in Fig.\ \ref{fig:Veff}. More details about $V_{\rm eff}$ and its properties are derived in Sec.\ \ref{sec:Veff}.

\subsection{Finite temperature corrections}

Thermal corrections to the potential decrease the value of the potential near the origin, but they have negligible effects at large field values, thus making the $x =0$ vacuum the true minimum above a certain critical temperature $T_c$. The thermal potential and its high-temperature expansion are given in the Appendix \ref{appA}.

\subsection{Interactions of fermions}
The fermionic part of the potential, following from the superpotential, contains the terms
\bea
    V_f &\supset& M_i \overline{N_i^C} N'_i + {\lambda'\over 2} \widetilde X \overline{N_i^C} N_i + Y_{i\alpha} \epsilon_{ab} \left(\sn_i \overline{\ho_u^{C,b}} P_L L_\alpha^a \right.\nn \\
    & +& \left.\slp^a_\alpha \overline{\ho_u^{C,b}} P_L N_i + H_u^b \overline{N_i^C} P_L L_\alpha^a  \right) + \mathrm{h.c}.
\eea
The second term on the first line violates lepton number conservation and gives rise to a Majorana mass $\mu=\lambda' \langle\widetilde X\rangle$ for $N_i$ after $\widetilde X$ acquires a VEV. In the limit $\mu\gg M$, $N_i$ and $N_i'$ split into two distinct eigenstates with masses $\sim \mu$ and
$M^2/\mu$ (see Eqs\ (\ref{eq:mNNp})).

 As was the case for scalar decays, the Yukawa couplings are the only source of CP violation for the fermionic decay channels of heavy neutrinos and sneutrinos.

\section{SUSY-breaking first-order phase transition}\label{sec:FOPT}

The leptogenesis mechanism presented in this paper (see Section \ref{sec:lepto}) does not require the SUSY-breaking phase transition to be first order. Nevertheless, FOPTs provide an interesting opportunity for probing high-energy physics by producing primordial black holes and a stochastic background of GWs. Such cosmological observables will be explored in Section \ref{sec:signatures} for the model presented in this paper. In this section, we present our methodology for finding and studying SUSY-breaking FOPTs and discuss their properties. 

\subsection{Bubble nucleation}

FOPTs proceed through bubble nucleations of the true vacuum. It is triggered by quantum tunnelling or thermal fluctuations, which can both be quantitatively described by an instanton or a bounce solution interpolating between the false and true vacua. In most cases of interest, the phase transition happens at high temperatures where thermal fluctuations are much more efficient than quantum tunnelling; hence we only describe the former here.

The nucleation rate, which can be derived from a semi-classical calculation, is given by \cite{Linde:1980tt}
\be
\Gamma(T) \simeq T^4\lp\frac{S_3}{2\pi T}\rp^{3/2}e^{-S_3/T}\,,
\ee
where $S_3$ is the O(3)-symmetric Euclidean action
\be\label{eq:action}
S_3(T) = 4\pi\!\!\int\! dr\, r^2\!\lc\frac{1}{2}\lp\frac{d\phi_i}{dr}\rp^2+V(\phi_i,T)\rc,
\ee
and $\phi_i$ denotes a scalar field whose VEV changes during the phase transition. The tunneling path $\phi_i(r)$ is found by requiring $S_3$ to be stationary, which leads to the equations of motion
\be
\frac{d^2\phi_i}{dr^2}+\frac{2}{r}\frac{d\phi_i}{dr}=\frac{\partial V}{\partial \phi_i}\,,
\ee
with the boundary conditions $\phi_i(r\rightarrow\infty)=\phi_i^{\rm false}$ and $\left.\frac{d\phi_i}{dr}\right\vert_{r=0} = 0$.

The phase transition begins at the nucleation temperature $T_n$, when there is an average of one bubble per Hubble volume \cite{Huber:2007vva}:
\be\label{eq:nucleation}
1 = \int_{T_n}^{T_c}\frac{dT}{T}\frac{\Gamma}{H^4} = \int_{T_n}^{T_c}dT\,\frac{T^3}{H^4}\lp\frac{S_3}{2\pi T}\rp^{3/2}e^{-S_3/T}\,.
\ee
$T_c$ is the critical temperature,  defined as the temperature where the true and false vacua are degenerate,
\be
\Delta V(T_c) = 0\,;
\ee
$\Delta V$ is the potential difference between the two vacua. To obtain an approximate solution of Eq.\ (\ref{eq:nucleation}), one can expand the argument of the exponential linearly in $T$, and fix $T=T_n$  elsewhere in the integrand; this is justified because the exponential varies much more rapidly than the other coefficients. This procedure yields the simpler condition
\bea\label{eq:Tn}
&&\left.\frac{S_3}{T}\right\vert_{T=T_n}\simeq 92.5+\frac{3}{2}\log\lp\frac{S_3}{2\pi T_n}\rp - 4\log\lp\frac{T_n}{10\ \mathrm{PeV}}\rp\qquad\nn\\
&&\quad-\log\lp\frac{\beta_H}{100(1-e^{-\beta_H(T_c/T_n-1)})}\rp - 2\log\lp\frac{g}{300}\rp\
\eea
where $g$ is the effective number of degrees of freedom and 
\be\label{eq:beta}
\beta_H =  \frac{\beta}{H} = T_n\left.\frac{d}{dT}\lp\frac{S_3}{T}\rp\right\vert_{T=T_n}
\ee
quantifies the inverse duration of the phase transition. Using the fact that the action is stationary under small variations of the tunneling path, one can write $\beta_H$ in terms of $\phi_i(r)$ and $S_3$ as
\be
\beta_H = 4\pi\!\!\int\!dr\, r^2\,\frac{\partial V}{\partial T}-\left.\frac{S_3}{T}\right\vert_{T=T_n},
\ee
which is numerically much more efficient than computing Eq.\ (\ref{eq:beta}) by finite difference.

Another important quantity for characterizing the strength of the phase transition is the ratio of vacuum energy released in the phase transition compared to the radiation energy, given by
\be\label{eq:alpha}
\alpha = \frac{1}{\rho_\gamma}\lp \Delta V-\frac{T_n}{4}\Delta\frac{dV}{dT}\rp,
\ee
where $\Delta V$ is the difference of potential energy between the true and false vacua and $\rho_\gamma = g\pi^2T_n^4/30$.  Larger $\alpha$ corresponds to more energy being released into the plasma, which yields a stronger phase transition. As we will see in Section \ref{sec:signatures}, this quantity, together with $\beta$, is important for determining the spectrum of GWs produced during the FOPT. In general,  the GW amplitude is enhanced at large $\alpha$ and small $\beta$.

\subsection{Effective bounce scalar potential}
\label{sec:Veff}

In order to study the dynamics of the phase transition, one must consider the scalar potential as a function of the fields whose VEVs vary across the bubble walls. As previously explained, these are the fields field $X$ and the four vectorlike messengers $\Phi$, $\Phi'$, $\overline\Phi$ and $\overline\Phi'$. The scalar potential appearing in the bounce action thereby reduces to 
\bea\label{eq:VB}
    V &=& \abs{F-\lambda \widetilde\Phi\widetilde{\overline\Phi}'}^2 + \left|m\right|^2\left(|{\widetilde\Phi}|^2+|{\widetilde{\overline\Phi}'}|^2\right)\nn\\
    &+&  \abs{\lambda \widetilde X \widetilde{\Phi} + m\widetilde{\Phi}'}^2 + \abs{\lambda \widetilde X \widetilde{\overline\Phi}' + m\widetilde{\overline\Phi}}^2\\
    &+& {g^2\over 2}\left({D\over g} + |\widetilde\Phi|^2 + |\widetilde\Phi'|^2 - |\widetilde{\overline\Phi}|^2  -|\widetilde{\overline{\Phi}}'|^2\right)^2 \nn\\
    &+& V_{CW}+V_T\,,
\eea
where $V_{CW}$ and $V_T$ are the 1-loop vacuum and thermal potentials, respectively (see Appendix \ref{appA} for more details).

Finding the bounce action for 10 real degrees of freedom is numerically expensive, so it is advantageous to simplify the scalar potential to reduce the effective number of degrees of freedom. This can be achieved with the methodology of Ref.\ \cite{Craig:2020jfv}. 
One first integrates out the $\Phi'$ and $\overline\Phi$ fields by solving for the $F$-terms of $\Phi$ and $\overline\Phi'$, which yields
\be\label{eq:Fterms}
\widetilde\Phi' = -\frac{\lambda}{m}\widetilde X\widetilde\Phi,\quad \widetilde{\overline\Phi}=-\frac{\lambda}{m}\widetilde X\widetilde{\overline\Phi'}\,.
\ee
Substituting Eqs.\ (\ref{eq:Fterms}) into Eq.\ \ref{eq:VB} gives the effective tunneling potential
\bea
V_{\rm eff} &=& \abs{F-\lambda \widetilde\Phi\widetilde{\overline\Phi}'}^2 + \left|m\right|^2\left(|{\widetilde\Phi}|^2+|{\widetilde{\overline\Phi}'}|^2\right)\nn\\
&+& \frac{g^2}{2}\lc \frac{D}{g}+\lp \abs{\widetilde\Phi}^2 - \abs{\widetilde{\overline\Phi}'}^2\rp\lp 1+\frac{\lambda^2}{m^2}\abs{\widetilde X}^2\rp\rc^2\quad\nn\\
&+& V_{CW}+V_T\,.
\eea
$V_{\rm eff}$ depends on just four fields: the magnitude of $x$, $\widetilde\Phi$, $\widetilde{\overline\Phi}'$, and the relative phase between the latter. Furthermore, the $F$-term of $X$ is minimized when this relative phase vanishes, so without loss of generality, we can choose the three fields to be real,  leaving three degrees of freedom. The bounce action Eq.\ (\ref{eq:action}) can then be rewritten in terms of the  fields $x=\sqrt{2}\vert\widetilde X\vert$, $\phi_1=\sqrt{2}\vert\widetilde\Phi\vert$ and $\phi_2=\sqrt{2}\vert\widetilde{\overline\Phi}'\vert$ as
\be
S_3^{\rm eff}(T) = 4\pi\!\!\int\!\! dr\, r^2\!\lc\frac{\Dot{x}^2}{2}+\frac{\Dot{\phi}_1^2}{2}+\frac{\Dot{\phi}_2^2}{2} + V_{\rm eff}(x,\phi_1,\phi_2;T)\rc .
\ee

To get accurate results, one must minimize the potential (\ref{eq:VB}) numerically to find the true vacuum. We  compute the bounce solution $(x,\,\phi_1,\,\phi_2)(r)$, 
and the tunneling action numerically using the package CosmoTransitions \cite{Wainwright:2011kj} which also calculates the nucleation temperature $T_n$ as described in the last subsection (see Ref.\ \cite{Craig:2020jfv} for an analytical estimate of the action).

One can get  analytic insight into the phase transition dynamics by estimating the position of the true vacuum. As argued in Section \ref{sec:framework}, the tree-level true vacuum is at $x\rightarrow\infty$, and it is rendered finite, though still large, by small radiative contributions to $V_{\rm eff}$. Consequently, Eq.\ (\ref{eq:runaway}) implies that $\langle\phi_1\rangle$ and $\langle\phi_2\rangle$ are small. Moreover, to minimize the $F$ and $D$-terms, $\phi_1$ must be suppressed with respect to $\phi_2$ by a factor of $gF/(\lambda D)$, which is assumed to be small. Therefore,  $\langle\phi_1\rangle$ is neglected in the following.

We integrate out $\phi_2$ by partially minimizing $V_{\rm eff}$ with respect to it, with the solution at large $x$
\bea
\langle\phi_2\rangle_{x\rightarrow\infty} &\cong& \frac{2m}{\lambda x}\sqrt{\frac{D}{g}}\,,\\
V_{\rm eff}(x\rightarrow\infty;T) &\cong& F^2+\frac{2m^4 D}{g\lambda^2x^2}+V_{CW}(x)+V_T(x;T)\,,\nn
\eea 
where we neglected the one-loop contributions to compute $\langle\phi_2\rangle$. As expected, the tree-level potential does not have a minimum at finite $x$. Therefore, one must estimate the one-loop contributions to obtain a finite VEV. We show in Appendix \ref{appA} that, for large $x$,
\bea
V_{CW}(x\rightarrow\infty) &\cong& \frac{3\lambda^2F^2}{8\pi^2}\log\lp\frac{\lambda^2x^2}{2m^2}\rp, \nn\\
V_T(x\rightarrow\infty;T) &\cong& \frac{3m^4T^2}{\lambda^2 x^2}\,,
\eea
neglecting constant terms that have no effect on the vacuum's position. Adding the one-loop corrections to the tree-level potential, one finds that the true vacuum is located at
\be
\langle x\rangle 
\cong \frac{4\pi m^2}{\sqrt{3}\lambda^2 F}\sqrt{\frac{D}{g}+\frac{3T^2}{2}}\,.
\label{xestimate}
\ee
$\langle x\rangle$ depends on negative powers of the coupling constants, confirming the hypothesis that $\langle x\rangle_{\rm true}$ is large when the interactions are weak. 

The fact that $x$ gains a large VEV has important consequences for the FOPT's dynamics. It implies that to get from one vacuum to the other, a large amount of kinetic energy must be expended, which increases the tunneling action. Therefore tunneling will only be possible at low temperature relative to $T_c$, which leads to greater supercooling and consequently a stronger FOPT. This has the potential to enhance GW production and  the probability of detection. In Section \ref{sec:signatures}, we will compute GW spectra for a range of parameters and show that this is indeed the case.

\section{Resonant leptogenesis}\label{sec:lepto}

\begin{figure*}[t]
		\centering
			\includegraphics[width=0.25\linewidth]{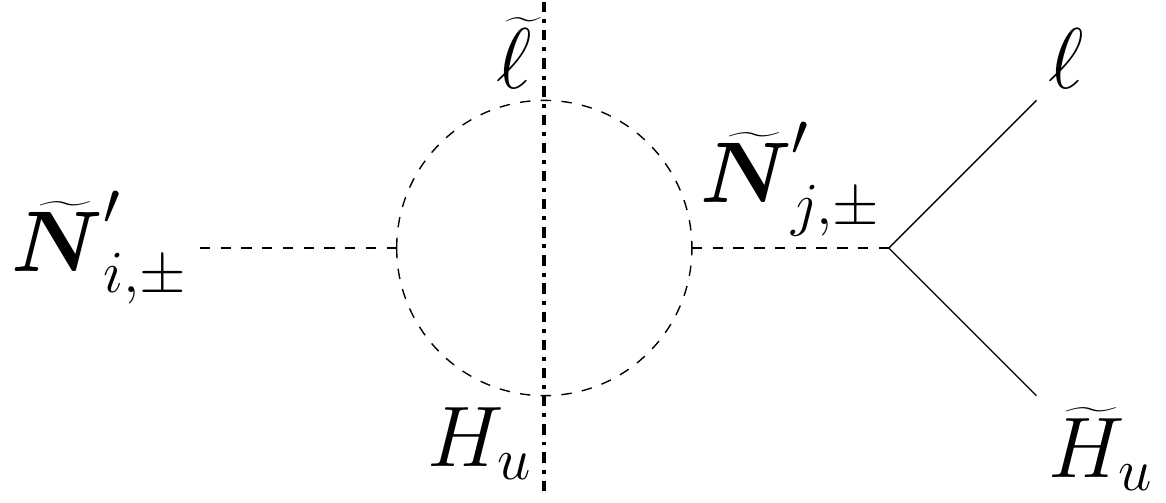}\includegraphics[width=0.25\linewidth]{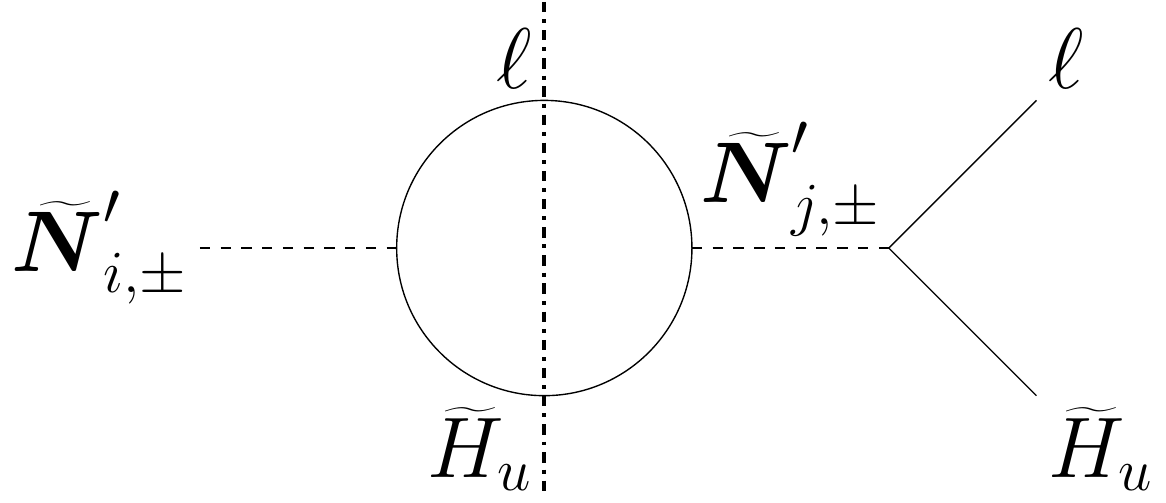}\includegraphics[width=0.25\linewidth]{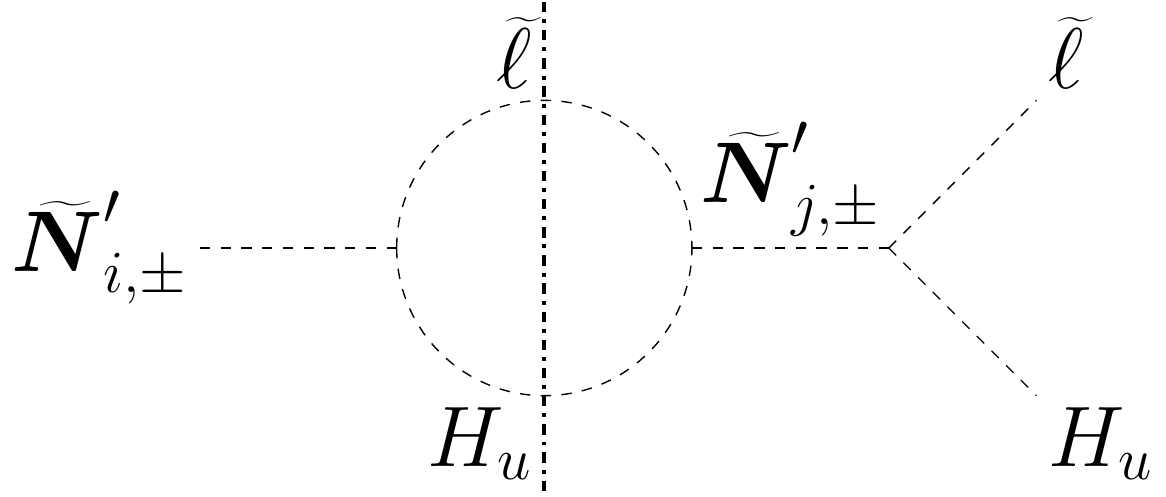}\includegraphics[width=0.25\linewidth]{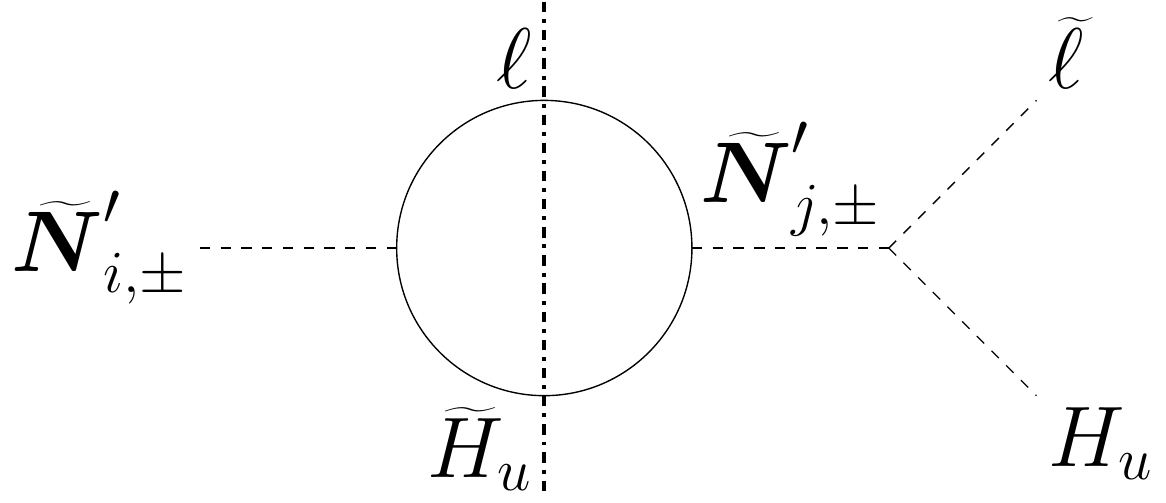}\\
			\includegraphics[width=0.25\linewidth]{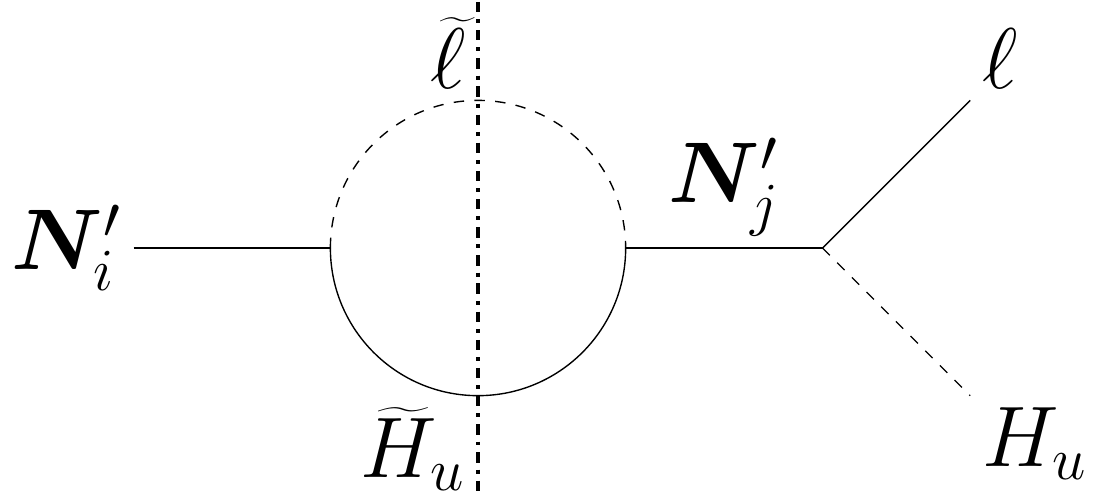}\includegraphics[width=0.25\linewidth]{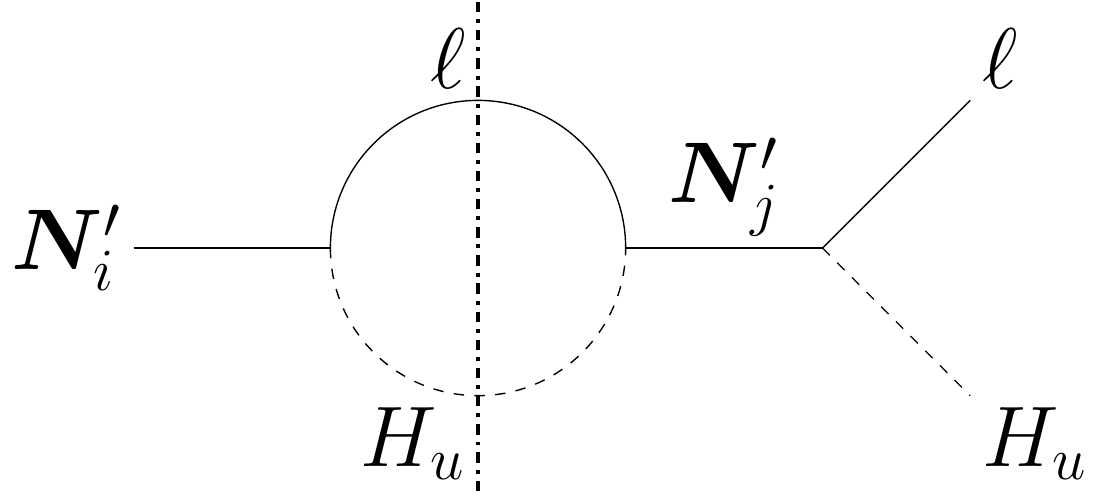}\includegraphics[width=0.25\linewidth]{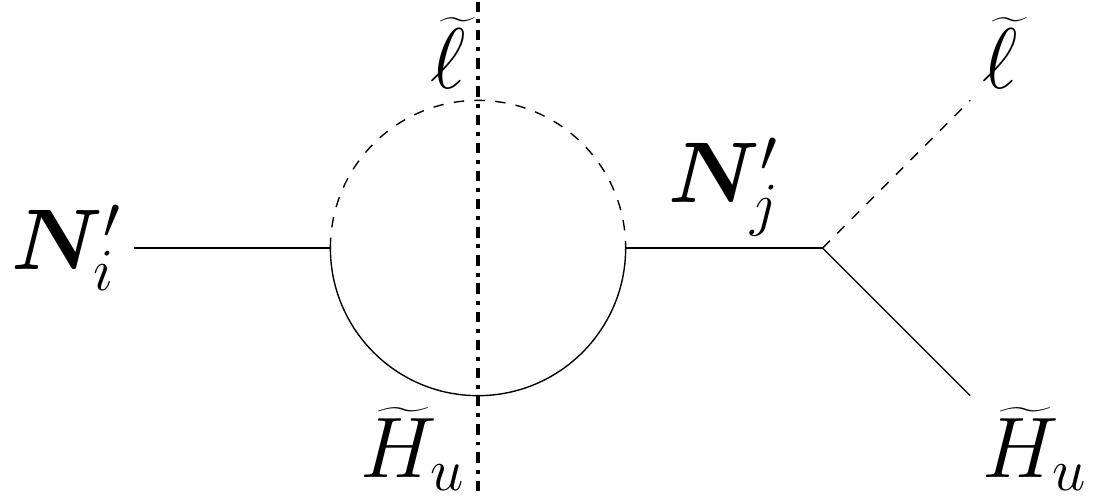}\includegraphics[width=0.25\linewidth]{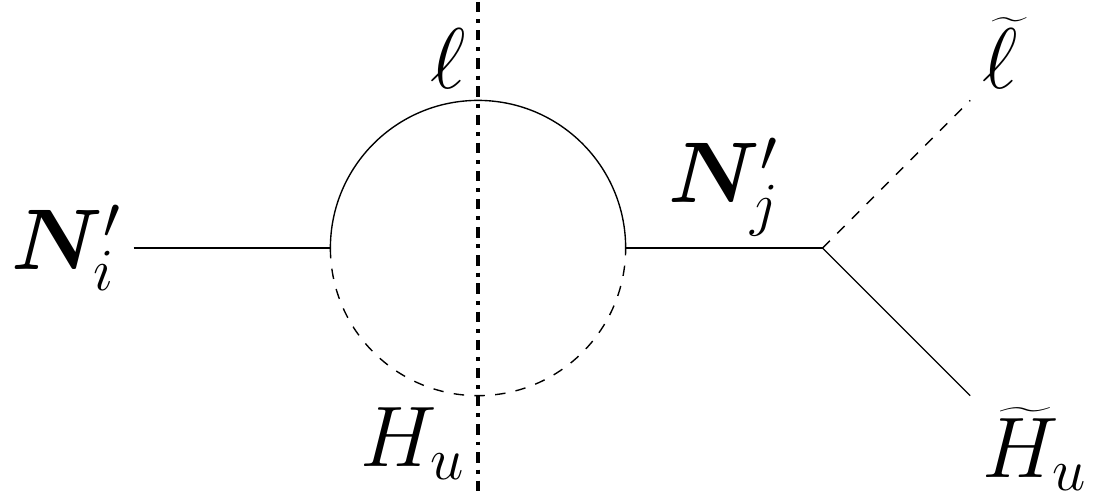}
		\caption{One-loop self energy diagrams for $\bm{\sn'}$  (top) and $\bm{N'}$ (bottom) decays. Dotted vertical lines indicate the intermediate states that go on-shell when evaluating the absorptive parts of the diagrams, in accordance with the Cutkosky rules. \cite{Cutkosky:1960sp}} \label{fig:loopdecay}
	\end{figure*}

Next we describe the mechanism of leptogenesis in the SUSY model, arising from the out-of-equilibrium decays of the sterile
neutrinos $N_i$, $N'_i$ and their superpartners $\sn_i$,
$\sn'_i$.
CP and lepton number violation in the decays generate a lepton asymmetry $Y_L^{\rm tot} = \sum_\alpha Y_{L_\alpha}+Y_{{\slp}_\alpha}$, where $Y_i=n_i/s$ is the comoving abundance of a given species. Assuming  $B-L$ is conserved, sphalerons partially convert the lepton asymmetry into a baryon asymmetry. In the MSSM, the conversion efficiency is \cite{Harvey:1990qw}
\be 
Y_B =\frac{n_B-n_{\overline{B}}}{s} = -\frac{8}{23} Y_L^{\rm tot}.
\ee 
BBN and CMB data constrain the baryon asymmetry of the universe to \cite{Workman:2022ynf,Planck:2018vyg}
\bea 
Y_B^{BBN} &&= (8.7\pm 0.5)\times 10^{-11}\nn\\
Y_B^{CMB}&& = (8.69\pm 0.06)\times 10^{-11} \label{eq:yb}
\eea 
at 95\%\,C.L.

In our model, all lepton-number-violating interactions are related to the breaking of SUSY via the nonvanishing $F$-term $F_X=F-\lambda\langle \widetilde{\Phi} \widetilde{\overline{\Phi}}'\rangle$ or the pseudomodulus $\widetilde{X}$,
\bea
-\mathcal{L}_{\not L} &=& - \frac{\lambda' F_X^*}{2} \sn_i \sn_i +  \frac{\lambda' \widetilde{X}}{2}\overline{N_i^C}N_i\\
&&+\lambda' \widetilde{X}\, \sn_i\lp M_i^* \sn_i'^* + \epsilon_{ab} Y_{i\alpha}^* \slp_\alpha^{a*} H_u^{b*} \rp.  \nn
\eea 
One may recognize the first term of this expression as a soft SUSY-breaking $B$-term, which mixes sneutrinos with their conjugates even before the phase transition, allowing their decays to violate the lepton number outside the bubble wall. When $\widetilde{X}$ gets a VEV, neutrinos acquire a Majorana mass $\mu = \lambda' \ev*{\widetilde{X}}$, opening their decays to lepton number violation as well.

A priori, all parameters appearing in the superpotential (\ref{superpot}) are complex. One can rephase the fields $N_i$, $N'_i$, $L_\alpha$ and $H_u$ to remove the complex phases in the mass matrix and in three Yukawa couplings, for example $Y_{1\alpha}$. If there are several right-handed neutrino flavors, in general one cannot simultaneously remove the phases of $Y_{1\alpha}$ and $Y_{j\alpha}$ ($j> 1$), allowing one to define a set of invariant CP-violating phases
\be 
\theta_{j\alpha} = \arg\lp Y_{j\alpha} Y^*_{1\alpha}\rp \qquad\text{(no sum on}\ \alpha;\  j>1),
\ee 
which can be assigned to $Y_{j\alpha}$. Hence at least two right-handed $N$ and $N'$ species are needed for leptogenesis.  Additional interactions such as  $A$-terms $A_i \sn_i \slp H_u$, which arise from RG running of SUSY-breaking interactions and which are essential to standard soft leptogenesis \cite{DAmbrosio:2003nfv,Fong:2011yx,Adhikari:2015ysa}, are not required and can be neglected if leptogenesis occurs at a scale close to $\sqrt{F}$. In the following we will consider two heavy neutrino flavors, so that only two SM neutrinos are massive.

There are two ways in which a lepton asymmetry can arise, depending on whether $\mu\ll M$ or $\mu\gg M$. In the former case, (s)neutrinos decay in the thermal bath, possibly even before the SUSY-breaking phase transition if $M\gg \sqrt{F}$ (in which case only sneutrinos produce a lepton asymmetry due to their $B$-term), and the CP asymmetry is resonantly enhanced by the small mass splitting between flavors. Strong washout is then required to achieve the observed baryon asymmetry. However, we have found that in this scenario, the SM neutrino masses induced at one loop are too small to match observations. 

Consequently, we will focus on the case $\mu \gg M$. After the phase transition, $N$ and $N'$ are weakly mixed with mixing angle $\sim M/\mu\ll1$, and we will refer to the corresponding mass eigenstates as
$\bm{N},\bm{N'}$.  Going to the basis where 
$M_{ij} = \left({M\atop -i M'}{iM'\atop M}\right)$ becomes diagonal with eigenvalues $M_i = M + s_1 M'$,
where $s_1=\pm 1$,
the four scalar masses are given by
\bea 
m^2_{\bm{\sn'}_{i,\pm}} &=& M_i^2 \lp \frac{M_i^2 +s_2 \lambda' F_X}{\mu^2}\rp \cong \frac{M_i^4}{\mu^2}, \nn\\
m^2_{\bm{\sn}_{i,\pm}} &=& \mu^2 + 2M_i^2 +s_2 \lambda' F_X\cong \mu^2, \label{eq:snmass}
\eea 
with $s_2 = \pm 1$.  In the approximation $m^2_{\bm{\sn'}_{i,\pm}} \cong M_i^4/\mu^2$, we assumed that $\lambda'F_X/\mu^2\ll M'/M$.
The fermionic mass eigenvalues are
\bea 
m_{\bm{N'}_i} &=&-{M_i^2\over \mu},  \nn\\
m_{\bm{N}_i} &=& \mu \lp 1+ {M_i^2\over \mu^2}\rp \cong \mu\,.\label{eq:mNNp}
\eea
It is useful to rewrite the Lagrangian interactions relevant for leptogenesis in the mass eigenbasis. Keeping only the leading terms,
one finds
\begin{align}
-\mathcal{L}_{s}&= \frac{\epsilon_{ab}Y_{i\alpha}}{\sqrt{2}} \times\nn \\
&\lb \bm{\sn'}_{i,-} \lp \frac{M_i^3}{\mu^2} \slp^a_\alpha H_u^b - \frac{M_i}{\mu} \overline{\ho^{C,b}_u}P_L L_\alpha^a \rp \right.\nn\\
&+ i\bm{\sn'}_{i,+} \lp \frac{M_i^3}{\mu^2} \slp^a_\alpha H_u^b + \frac{M_i}{\mu} \overline{\ho^{C,b}_u}P_L L_\alpha^a \rp \nn\\
&+\bm{\sn}_{i,-} \lp \mu \slp^a_\alpha H_u^b +  \overline{\ho^{C,b}_u}P_L L_\alpha^a \rp \nn\\ 
&\left.+ i\bm{\sn}_{i,+} \lp \mu \slp^a_\alpha H_u^b -  \overline{\ho^{C,b}_u}P_L L_\alpha^a \rp \rb + \rm{h.c.} 
\end{align}
for sneutrinos and 
\begin{align}
-\mathcal{L}_{f}&= \epsilon_{ab}Y_{i\alpha} \times\nn \\
&\lb \bm{\overline{N_i^{\prime C}}} \lp- \frac{M_i}{\mu} \slp_\alpha^a P_L \ho_u^b   - \frac{M_i}{\mu} H_u^b P_L L_\alpha^a\rp \right.\nn\\
& + \bm{\overline{N_i^{C}}} \lp\slp_\alpha^a P_L \ho_u^b   +  H_u^b P_L L_\alpha^a\rp \bigg\} + \rm{h.c.} \label{eq:lagN}
\end{align}
for neutrinos. 

When $\widetilde{X}$ acquires a VEV, the $\bm{N}$ states become heavy and 
quickly decay, giving rise to an initial lepton asymmetry, due to the CP-violating phases from the Yukawa couplings. In contrast, $\bm{N'}$ is typically lighter than the SUSY-breaking scale.
 \ As we will show, because of its small mass and strong interaction with the thermal bath, $\bm{N'}$ remains in equilibrium and its inverse decays exponentially wash out the lepton asymmetry produced by $\bm{N}$ decays. Therefore, we can ignore the $\bm{N}$ contribution to the asymmetry and focus on that of $\bm{N'}$.

When $T$ drops below $m_{\bm{N'}}\cong M^2/\mu$, the $\bm{N'}$ states fall out of equilibrium and a net lepton asymmetry results. Within the MFV hypothesis, CP violation in (s)neutrino decays comes from the one-loop mixing between $\bm{N'}$ ($\bm{\sn'}$) flavors. The one-loop diagrams are illustrated in Fig.\ \ref{fig:loopdecay}. The detailled calculation of the CP asymmetry will be presented in section \ref{sec:cp}. 

Ignoring corrections to the mass eigenvalues, $\bm{N'_i}$ and $\bm{\sn'_i}$ states have the same decay rates into leptons, sleptons and their conjugates. The total decay rate is given by
\be \label{eq:decaywidth}
\Gamma_i = \frac{1}{4\pi} \left( YY^\dagger\right)_{ii} \left(\frac{M^4}{\mu^3} \right).
\ee

\begin{figure*}[t]
    \centerline{
    \includegraphics[scale=0.4]{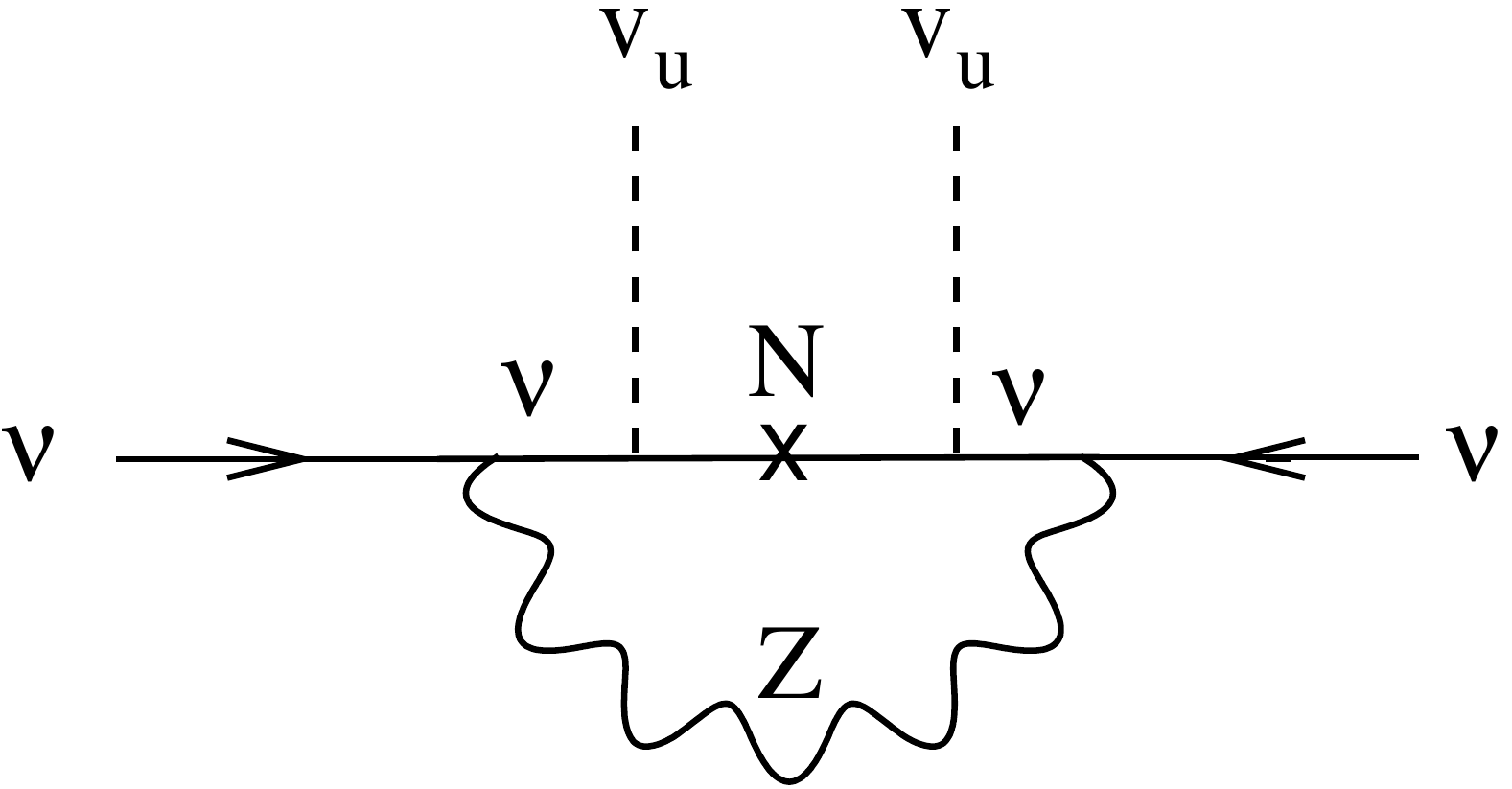}\hfil
      \raisebox{0.25cm}{ \includegraphics[scale=0.4]{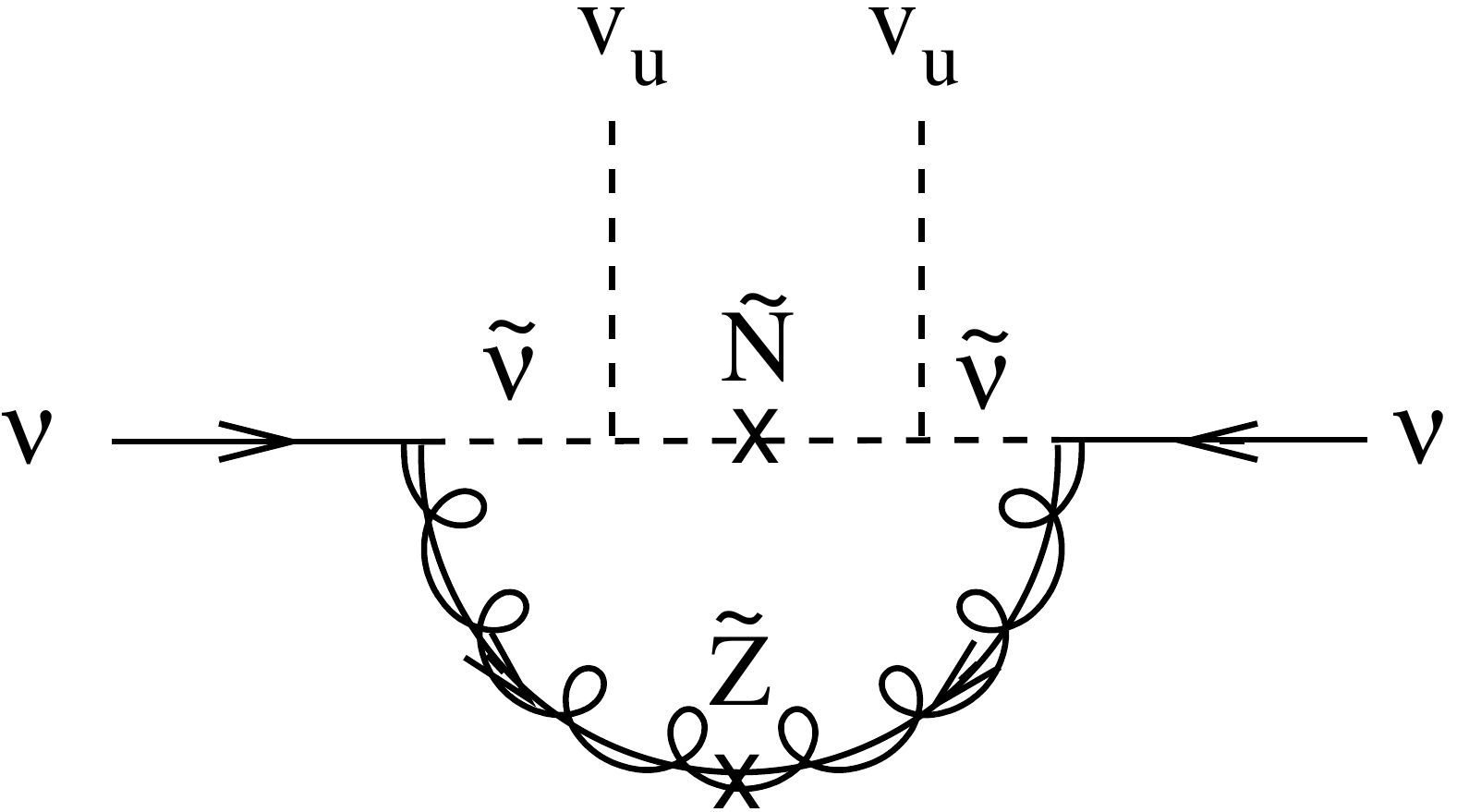}}}
    \caption{Mass generation of the light neutrinos.}
    \label{fig:nuloop}
\end{figure*}
\subsection{Loop-induced neutrino masses}

To fully determine the parameters consistent with successful leptogenesis, we must
relate the Yukawa couplings to the observed light neutrino masses.
The mass matrix in the basis $(\nu,\,N',\,N)$ takes the form
\be
    \left(\begin{array}{ccc}0 & 0 & m_D\\
       0  & 0 & M\\
        m_D^T & M & \mu \end{array}\right)\,,
\ee
where $m_{D,i\alpha} = Y_{i\alpha}\langle H_u\rangle \equiv Y_{i\alpha} v_u$,  $\mu_{ij} = \lambda'\langle\widetilde{X}\rangle\delta_{ij}$ and $M$
denotes the matrix $M_{ij}$ from Eq.\ (\ref{superpot}).  We recall our assumption that $|\mu|\gg |M|$.  The lightest neutrinos remain massless with this matrix.  However at one loop, a  direct Majorana mass  $m_\nu$ is generated for the $\nu$ states by the diagrams shown in Fig.\ \ref{fig:nuloop}.  If $M\gg m_D$, mixing between $\nu$ and $N'$ can be neglected, and the light neutrino mass matrix is given by 
\bea\label{eq:numass}
    m_\nu &=& { g_2^2\,Y^T  Y\,(v_u^2/\mu)\over 8\pi^2\cos^2\theta_W}\left({\ln(\mu/m_Z)\over
        1-m_Z^2/\mu^2} +{\lambda' F_X\, \over 16\,\mu\,m_{\tilde Z}}f(s_m)  \right)\,,\nn\\
    f(x) &=& x^2(1-x^2 + x^2\ln(x^2))\over (1-x^2)^2\,,
\eea
where $g_2$ is the SU(2) gauge coupling, $m_{\tilde Z}$ is the Zino mass, 
\be \label{eq:mzino}
m_{\tilde Z} =N_m \frac{g_2^2}{(4\pi)^2} \frac{F_X}{\langle\widetilde X\rangle}s_m C_{RG} \, ,
\ee 
including gaugino screening by the factor $s_m$, $N_m=2$ sets of
messenger fields and RG running correction $C_{RG}$ as in Eq.~(\ref{eq:mgluino}).
We have approximated $m_{\tilde Z} \cong m_{\tilde\nu}\, s_m$, and that $M$ is proportional to the unit matrix in the $N$-$N'$ flavor space. 
We find that $m_\nu$ is dominated by the first diagram, with the second making a correction of $\sim 8\,\%$ for a screening factor of $s_m\sim 0.1$.

In addition to the previous assumptions, and the near-degeneracy of the heavy RH neutrino $\bm{N_i}$ we consider the SM neutrinos to be hierarchical. To relate light neutrino masses with the parameters of our model, we introduce a Casas-Ibarra parametrization of the Yukawa matrix \cite{Casas:2001sr,Ibarra:2003up,Garbrecht:2014aga}.
Because our model only contains two sterile neutrino flavors,  the lightest neutrino is exactly massless. We will consider the normal (NH) and inverted (IH) hierarchies, for which $m_1\cong 0$ and $m_3\cong 0$ respectively.   One   obtains
\bea
Y_{i\alpha} = \sqrt{C\frac{ \mu}{v_u^2}}
		\,\mathcal{R}\, D_{\sqrt{m_\nu}}\, U_\nu^\dagger
\label{eq:Yias}
\eea
where we have regrouped all numerical coefficients into $C\approx 9.7$, $U_\nu$ is the PMNS matrix and $D_{\sqrt{m_\nu}}$ is the diagonal matrix  
\bea
D_{\sqrt{m_\nu}} &&\,= {\rm diag} (0,\sqrt{m_2},\sqrt{m_3})\qquad {\rm (NH)},\nn\\
D_{\sqrt{m_\nu}} &&\,= {\rm diag} (\sqrt{m_1},\sqrt{m_2},0)\qquad {\rm (IH)}.
\eea 
The $2\times 3$ matrix $\mathcal{R}$ contains a $2\times2$ complex orthogonal submatrix,
\bea
\mathcal{R} &&\,= \lp\begin{array}{ccc}
	\phantom{-}0&\phantom{-} \cos \hat\varrho & \sin \hat\varrho  \\
	\phantom{-}0&-\sin \hat\varrho & \cos \hat\varrho
\end{array}\rp,\qquad {\rm (NH)}\nn\\
\mathcal{R} &&\,= \lp\begin{array}{ccc}
	\phantom{-} \cos \hat\varrho & \sin \hat\varrho &\phantom{-}0 \\
	-\sin \hat\varrho & \cos \hat\varrho&\phantom{-}0
\end{array}\rp,\qquad {\rm (IH)}, \label{eq:rmatrix}
\eea 
where $\hat\varrho\equiv a+ib$ is a complex parameter.  Using 
 Eq.~(\ref{eq:numass}), one can estimate the numerical values of the Yukawa couplings required to yield neutrino masses in agreement with observations,
\bea 
Y^TY &=& 1.6\times10^{-6} \lp \frac{\mu}{100\ \rm PeV} \rp \nn\\
&\times&\lp\frac{m_\nu}{0.05\ \rm eV}\rp\lp\frac{1}{\sin^2 \beta} \rp  
\eea 
where  $\tan\beta = v_u/v_d$ and we used $v_u^2+v_d^2 = (174\ \rm{GeV})^2$.

The matrix $YY^\dagger$ will be central in our analysis of leptogenesis, so it is convenient to express it using the Casas-Ibarra parametrization, 
\bea  \label{yydagger}
YY^\dagger &\equiv& C\frac{\mu}{2v_u^2} \, \mathcal{M}
\\ \nn &=&C\frac{\mu}{2v_u^2} \lp\begin{array}{cc}
	\phantom{-}\delta\,\c_{2a}+ \sigma\,\ch_{2b}  & -\delta\, \s_{2a} + i \sigma\,\sh_{2b} \\  
	-\delta\, \s_{2a} - i \sigma\,\s_{2b} & -\delta\, \c_{2a}+ \sigma \,\ch_{2b} \\
\end{array}\rp,
\eea 
where $\sigma = m_2+m_3$ ($m_1+m_2)$ and $\delta = m_2-m_3$ ($m_1-m_2)$ for NH (IH). In this expression, $a$ and $b$ are the real and imaginary parts of the complex angle $\hat\varrho$ in Eq.~(\ref{eq:rmatrix}), and $\c_x=\cos x,\s_x=\sin x, \ch_x=\cosh x,\sh_x=\sinh x$. Note that the PMNS matrix doesn't enter this expression and therefore has no direct impact on leptogenesis.

\subsection{CP asymmetry}
\label{sec:cp}

Because of the tiny mass splitting $\sim M'$ between sterile neutrino flavors, one-loop flavor mixing is resonantly enhanced. This makes the self-energy diagrams of Fig. \ref{fig:loopdecay} the dominant contribution to CP asymmetry in $\bm{N'}$ and $\bm{\sn'}$ decays, which is often referred to as $\varepsilon$-type CP violation. Vertex diagrams, or $\varepsilon'$-type CP violation, can therefore be neglected in our model.  \cite{Fong:2011yx,Adhikari:2015ysa,Pilaftsis:2003gt,Pilaftsis:1997jf}

Two types of sneutrino mixing can lead to CP violation: $\bm{\sn'_{i,\pm}}$ with $\bm{\sn'_{j,\pm}}$ (same sign), and $\bm{\sn'_{i,\pm}}$ with $\bm{\sn'_{j,\mp}}$ (opposite sign). In the former case, the mass squared difference is of order (\emph{cf.} Eq.~(\ref{eq:snmass})
\be \label{eq:deltam2}
\delta m_{ij}^2 \cong 8 M^2 \lp M \over \mu\rp^2\lp M'\over M\rp,
\ee 
which is much smaller than the splitting of the second mixing, $\sim \lambda'F_X (M/\mu)^2$. Therefore, the second mixing is not as resonant as the first one, and we can ignore CP violation coming from the latter. 

Using the resummation approach for unstable particle propagators described in \cite{DAmbrosio:2003nfv,Pilaftsis:1997jf,Fong:2011yx,Pilaftsis:2003gt}, the one-loop amplitude for the sneutrino decays  $\bm{\sn_i'}\to a_\alpha$, with $a_\alpha = \slp_\alpha H_u$ or $L_\alpha \ho_u$ is	
\be 
\hat{\mathcal{A}}^{a_\alpha}_i = A_i^{a_\alpha} - i \sum_{j\neq i} A_j^{a_\alpha} \frac{ \widetilde\Sigma^\ab_{ji}}{p^2-M_j^2+i\widetilde\Sigma^\ab_{jj}}
\ee
In this expression, $A_i^{a_\alpha}$ is the tree-level amplitude of the $\bm{\sn_i'}\to a_\alpha$ decay and $\widetilde\Sigma_{ji}^\ab$ is the absorptive part of the $\bm{\sn_i'}\to \bm{\sn_j'}$ self-energy, which we evaluated with the Cutkosky rules \cite{Cutkosky:1960sp}.

For neutrino decays $\bm{N_i'}\to b_\alpha$, where $b_\alpha = L_\alpha H_u$ or $\slp_\alpha \ho_u$, the one-loop amplitude is given by a similar expression,
\be 
\hat{\mathcal{A}}^{a_\alpha}_i = \overline{u}_{b_\alpha} P_R \left\lbrace h_{i\alpha} - i \sum_{j\neq i} \frac{h_{j\alpha}}{\slashed{p}-M_j + i \Sigma^\ab_{jj}} \Sigma_{ji}^\ab\right\rbrace u_{\bm{N'_i}}.
\ee 
Here, $\Sigma_{ji}^\ab$ is the absorptive part of the $\bm{N_i'}\to \bm{N_j'}$ self-energy and $h_{i\alpha}$ is the tree-level coupling between $\bm{N'_i}$ and the final state $b_\alpha$, \emph{cf.} Eq.~(\ref{eq:lagN}).

One can define the CP asymmetry parameters as
\be 
\epsilon_i = \frac{\sum_{a_\alpha,\alpha} \Gamma(\bm{N'_i}\to a_\alpha) - \Gamma(\bm{N'_i}\to a_\alpha^*) }{\sum_{a_\alpha,\alpha} \Gamma(\bm{N'_i}\to a_\alpha) + \Gamma(\bm{N'_i}\to a_\alpha^*)}
\ee
for neutrinos and
\be 
\widetilde\epsilon_i = \frac{\sum_{b_\alpha,\alpha} \Gamma(\bm{\sn'_i}\to b_\alpha) - \Gamma(\bm{\sn'_i}\to b_\alpha^*) }{\sum_{b_\alpha,\alpha} \Gamma(\bm{\sn'_i}\to b_\alpha) + \Gamma(\bm{\sn'_i}\to b_\alpha^*)}
\ee 
for sneutrinos. Here, $a_\alpha^*$ and $b_\alpha^*$ are the CP conjugates of the final states $a_\alpha$, $b_\alpha$.

At one-loop order, we find that neutrinos and sneutrinos have  the same CP asymmetry:
\be \label{eq:epsilon}
\epsilon_i =\widetilde{\epsilon}_i =\frac{1}{2} \sum_{i\neq j} \frac{\Im\lbr(YY^\dagger)^2_{ij}\rbr}{(YY^\dagger)_{ii}(YY^\dagger)_{jj}} \frac{(m_i\Gamma_j)\, (\delta m^2_{ij}) }{(\delta m^2_{ij})^2 + (m_i\Gamma_j)^2}\,.
\ee 
This expression is the same as in resonant leptogenesis \cite{Pilaftsis:2003gt,Garbrecht:2014aga}. Significantly, because all decay channels of a given state contribute with the same sign to the CP asymmetry, thermal factors of the decay product cancel out in the expression for $\epsilon_i$, which implies the CP asymmetry survives even in the $T\to 0$ limit. This is to be contrasted with the standard case of soft leptogenesis where bosonic and fermionic decay channels have opposite CP asymmetry and leptogenesis requires thermal effects to avoid exact cancellation between them \cite{DAmbrosio:2003nfv,Fong:2011yx,Adhikari:2015ysa}.

In the absence of SUSY breaking, the resonance condition $m_i \Gamma_j \sim \delta m_{ij}^2$ would be satisfied assuming $Y^2\sim M'/M$, as can by seen by comparing Eqs.~(\ref{eq:snmass}), (\ref{eq:mNNp}), (\ref{eq:decaywidth}) and (\ref{eq:deltam2}). In that case, the fraction $(m_i\Gamma_j) \delta m^2_{ij}/[(\delta m^2_{ij})^2
+(m_j\Gamma_j)^2]$ is maximized, leading to $\epsilon_i\sim 1$. This would be an example of resonant leptogenesis. 

However, in our model the breaking of SUSY suppresses the decay width $m_i\Gamma_j$ by $(M/\mu)^4$ ({\it cf.}, Eq.~(\ref{eq:decaywidth})) while $\delta m^2$ is suppressed by $(M/\mu)^2$ only  ({\it cf.}, Eq.~(\ref{eq:deltam2})).

 With  $M'/M = 10^{-7}$, we find that $\delta m_{ij}\gg m_i \Gamma_j$ and the resonant enhancement is not maximal, but it is still sufficient to yield successful leptogenesis at the PeV scale, as we will show. We can rewrite Eq.~(\ref{eq:epsilon}) in terms of the leptogenesis scale, $M^2/\mu$,
\be \epsilon_i \simeq \frac{C}{64\pi\, v_u^2} \lp{M\over M'}\rp  \lp \frac{M^2}{\mu}\rp \sum_{j\neq i}  \frac{\abs{\mathcal{M}}^2_{ij}\sin(2 \phi_{ij})}{(\mathcal{M})_{ii}},\label{eq:epsilon2}
\ee
where $\phi_{ij}=\arg(\mathcal{M}_{ij})$. We recall that $C\approx 9.7$ and $\mathcal{M}$ is the matrix introduced in the Casas-Ibarra parametrization of $YY^\dagger$, Eq.~(\ref{yydagger}). Assuming $M^2/\mu\approx 1$ PeV, $\mathcal{M}\approx 0.05$ eV and $M'/M=10^{-7}$, and barring a strong hierarchy between on- and off-diagonal entries of the matrix $\mathcal M$ (or equivalently, of $YY^\dagger$), one could therefore expect the CP asymmetry to be of order $\epsilon_i\sim 10^{-3}$. Although this is not as large as what standard resonant leptogenesis models can achieve, this is sufficient to produce the observed baryon asymmetry as we will now show.

\subsection{Evolution of lepton asymmetry}
\begin{figure*}[ht]
    \centering
    \includegraphics[width=0.45\linewidth]{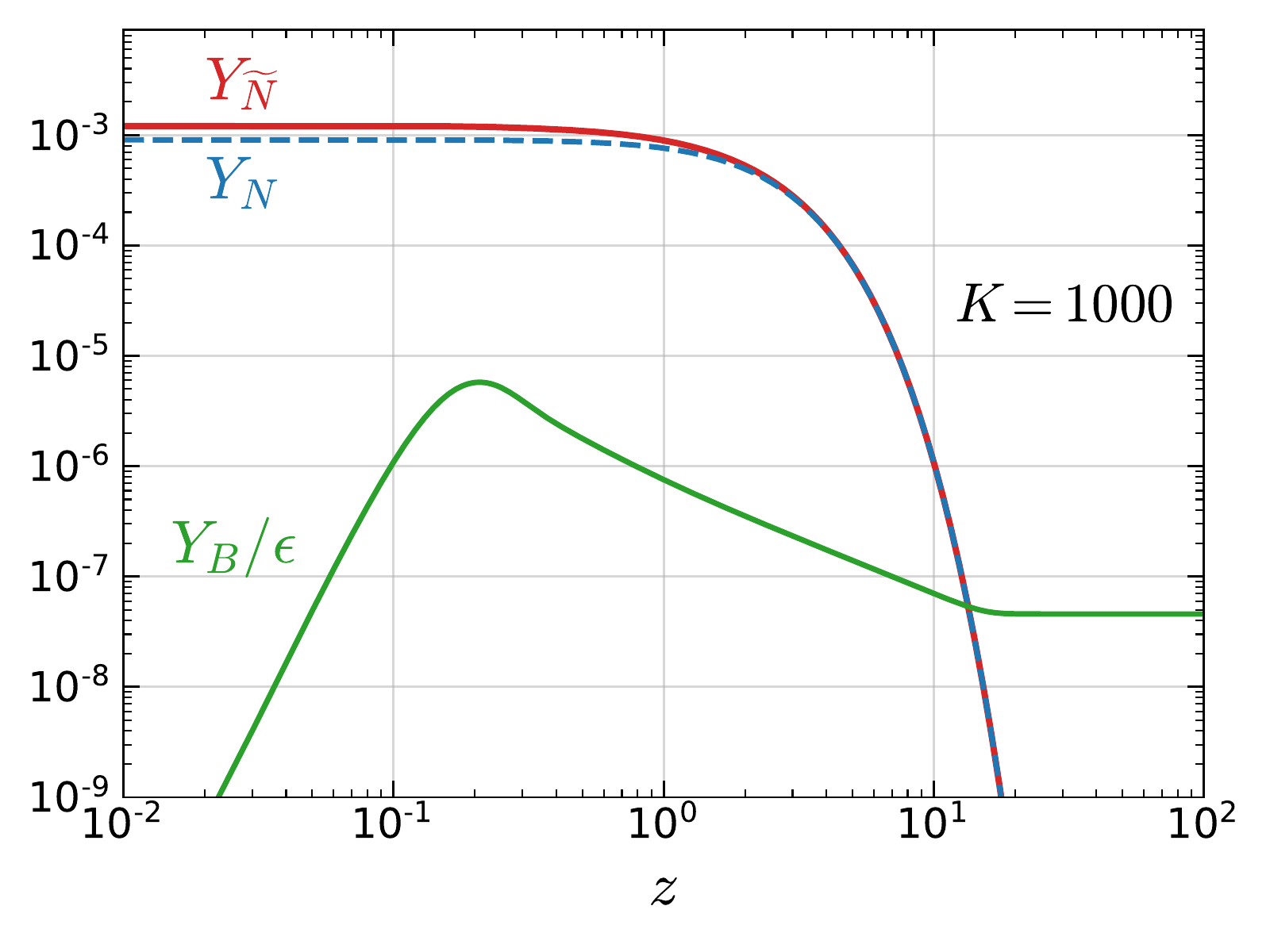}\includegraphics[width=0.45\linewidth]{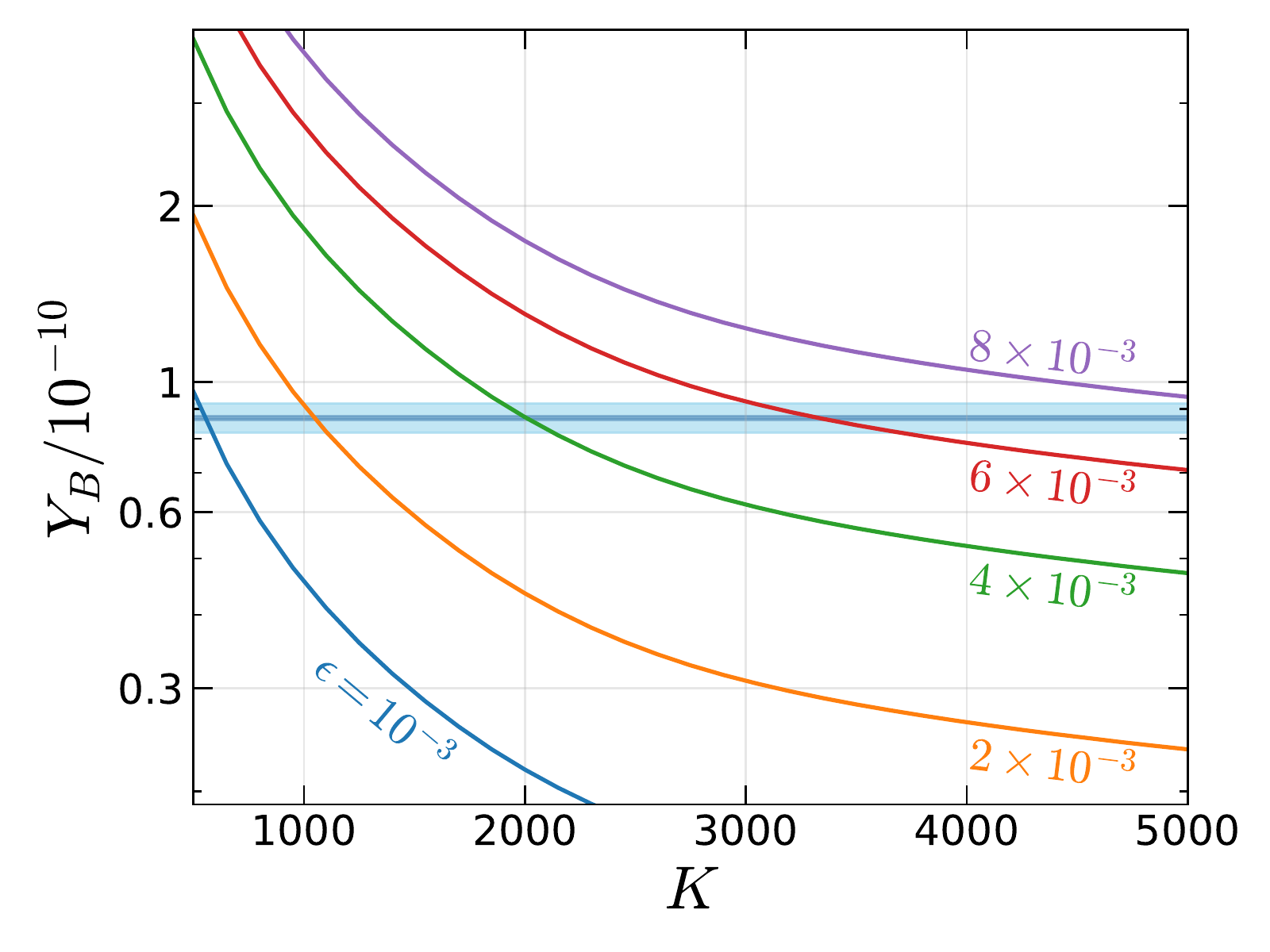}\\[-3mm]
    \textbf{(a)}\hspace{0.48\linewidth}\textbf{(b)}\hspace{0.35\linewidth}
    \caption{(a) Evolution of the (s)neutrino abundance $Y_N$ ($Y_\sn$) and the baryon asymmetry as a function of $z=m_i/T=(M^2/\mu)/T$ for a benchmark model with washout parameter $K=1000$. (b) Final baryon asymmetry for varying washout parameter $K$ and CP asymmetry $\epsilon$. The light (dark) blue horizontal band illustrates the 95\,\% C.L. limits set by BBN (CMB) data., \textit{cf.} Eq.~(\ref{eq:yb}).}
    \label{fig:yb}
\end{figure*}
The initial asymmetry generated by the decays is partially washed out by scattering and inverse decay processes.  The strength of the latter is characterized by a washout parameter, defined as
\be 
K_i = \frac{\Gamma_i}{H_i}\,,
\ee 
where $\Gamma_i$ is given in Eq.~(\ref{eq:decaywidth}) and $H_i = 1.66\sqrt{g_*} \,m_i^2/M_{\rm pl}$ is the Hubble rate evaluated at $T = m_i$, with Planck mass $M_{\rm pl}=1.22\times 10^{19}$ GeV. Numerically estimating the value of $K_i$ yields
\be \label{eq:washout}
K_i = 1000 \lp \frac{(YY^\dagger)_{ii}}{1.6\times 10^{-6}} \rp  \lp\frac{100\ {\rm PeV}}{\mu} \rp \lp\frac{230}{g_*} \rp^{1/2},  
\ee 
which is in the strong washout regime,  $K_i\gg 1$. 

In the strong washout regime, inverse decays of right-handed neutrinos remain in equilibrium down to temperatures well below their masses. This implies significant suppression of the final lepton asymmetry, but it also allows for several simplifying assumptions:
\begin{itemize}
    \item the final lepton asymmetry is independent of the initial right-handed neutrino abundances because they rapidly come into thermal equilibrium;
    \item any lepton asymmetry coming from the decays of heavy $\bm{N}$ and $\bm{\sn}$ (whose mass is $\mu \gg M,\sqrt{F}$) is exponentially suppressed and can be ignored relative to those of the lighter $\bm{N'}$ and $\bm{\sn'}$;
    \item thermal effects and CP violation from $2\to 2$ scattering processes can be ignored, as those are significant only at large temperature  \cite{Buchmuller:2004nz}.
\end{itemize}
The coupled Boltzmann equations for the evolution of the 
right-handed neutrino, sneutrino and lepton number abundances, $Y_X = n_X/s$, are
\bea 
\frac{dY_{\bm{N'_i}}}{dz} &&= - K_i z \frac{\mathcal{K}_1(z)}{\mathcal{K}_2(z)} \lp Y_{\bm{N'_i}}- Y_{\bm{N'_i}}^{\rm eq} \rp \\
\frac{dY_{\bm \sn'_i}}{dz} &&= - K_i z \frac{\mathcal{K}_1(z)}{\mathcal{K}_2(z)} \lp Y_{\bm \sn'_i}- Y_{\bm \sn'_i}^{\rm eq} \rp \\
\frac{dY_L}{dz} &&= \sum_i\lbr \epsilon_i K_i z \frac{\mathcal{K}_1(z)}{\mathcal{K}_2(z)} \lp Y_{\bm{N'_i}}- Y_{\bm{N'_i}}^{\rm eq} + Y_{\bm \sn'_i} - Y_{\bm \sn'_i}^{\rm eq} \rp \right. \nn\\
 &&\left. -2\ \frac{K_i z^3}{4} \mathcal{K}_1(z) Y_L \rbr
\eea 
where $z=m/T=(M^2/\mu)/T$, indicating our approximation that all states have the same mass, ${\cal K}_i(x)$ are the modified Bessel functions of the second kind, and the equilibrium abundances in the Maxwell-Boltzmann approximation are given by
$Y^{\rm eq} = 45/(4 \pi^4 g_*) z^2 \mathcal{K}_2$. 

In Fig.\ \ref{fig:yb}(a) we show the numerical solutions of
the Boltzmann equations assuming the two right handed neutrino flavors have the same washout parameter $K_i=1000$ and the same CP asymmetry $\epsilon_i$. The initial conditions we used were
$Y_{\bm N'_i} = Y_{\bm \sn'_i}=Y^{\rm eq}$ and $Y_L =10^{-12}$, from the initial decays of the heavy states.  The final values relevant for the observed baryon asymmetry are however quite insensitive to the initial conditions, which is characteristic of the strong washout regime.  In Fig.\ \ref{fig:yb}(b), the resulting baryon asymmetry as a function of $K$ is plotted for several values of $\epsilon$. Numerically, we obtain that the solution $Y_L$ scales as 
\be 
Y_L \sim 10^{-4} \frac{\epsilon}{K}
\ee 
up to a logarithmic dependence on $K$, which agrees with the analytical estimate found in Ref. \cite{Buchmuller:2004nz} for the strong washout regime. These estimates show that we can obtain the observed asymmetry with $\epsilon\sim 10^{-3}$.  
\begin{figure*}[t]
    \centering
    \includegraphics[width=0.5\linewidth]{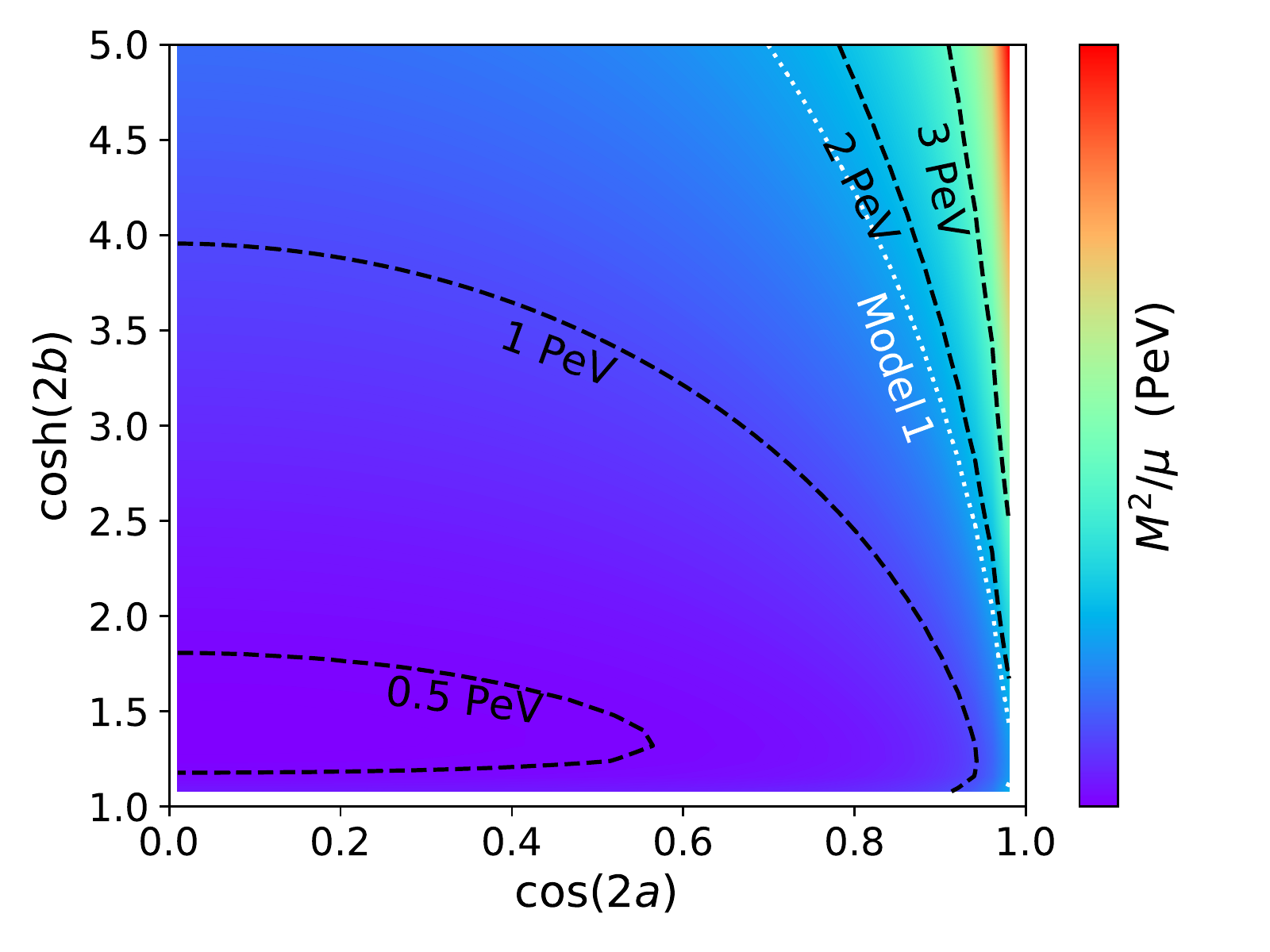}\includegraphics[width=0.5\linewidth]{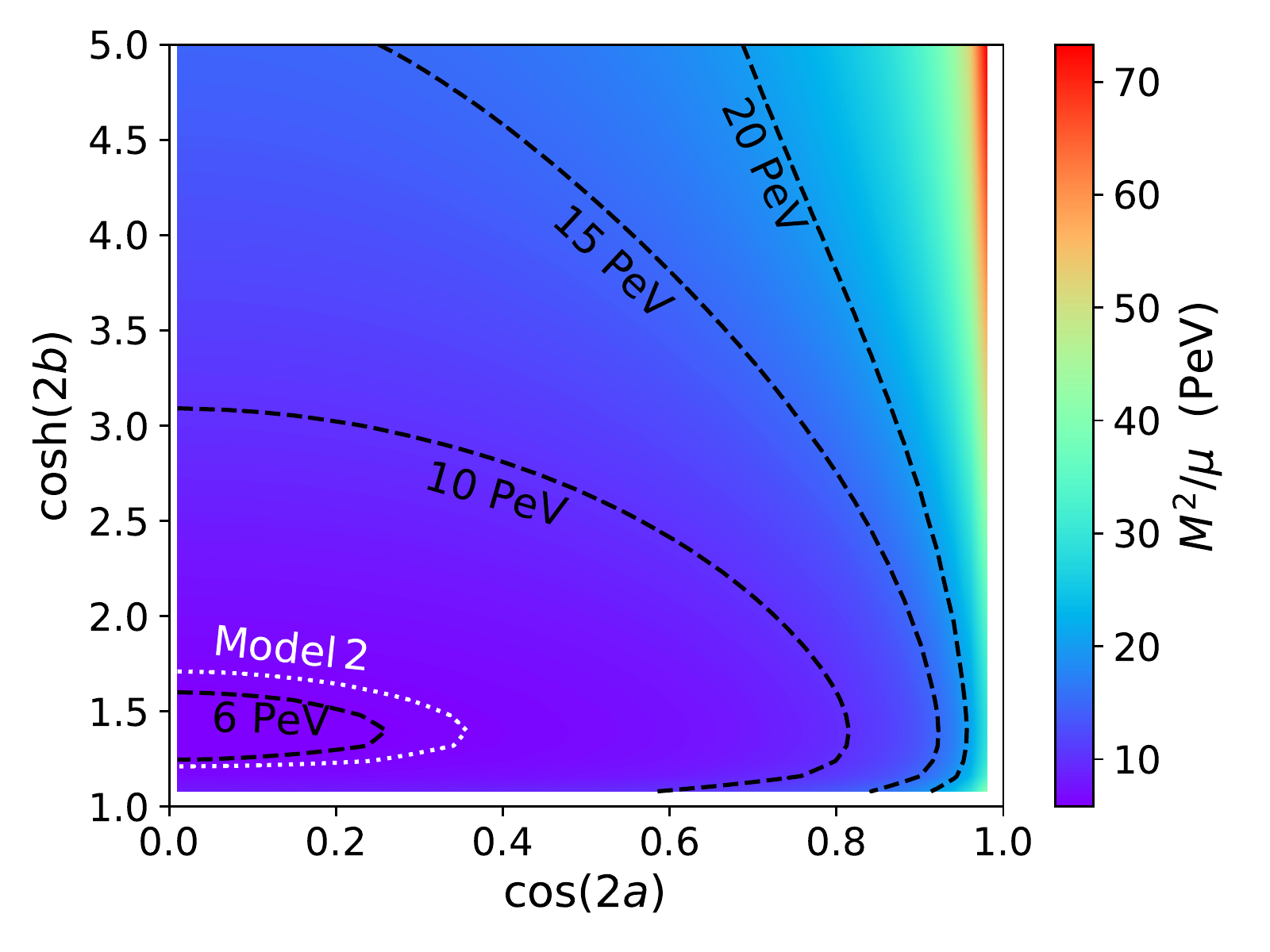}\\[-3mm]
    \textbf{(a)}\hspace{0.48\linewidth}\textbf{(b)}\hspace{0.35\linewidth}
    \caption{Contour plot of the leptogenesis scale $M^2/\mu$ that yields the observed baryon asymmetry for a given complex angle $\hat \varrho=a+ib$ in the $\mathcal{R}$ matrix of Eq. \ref{eq:rmatrix}. The left and right panels show scenarios with normal and inverted light neutrino hierarchy, respectively. Black dashed curves show the contour lines of some benchmark values. White dotted curves show parameters corresponding to models 1 (left) and 2 (right) of Table \ref{tab2}.}
    \label{fig:yl-scans}
\end{figure*}

Combining the Casas-Ibarra parametrization of $YY^\dagger$ (\ref{yydagger}) with the expression for the washout parameter (\ref{eq:washout}) and the CP asymmetry (\ref{eq:epsilon2}), one finds that the final lepton asymmetry depends on a handful of parameters, namely the light neutrino masses, the leptogenesis scale $M^2/\mu$, the mass splitting parameter $M'/M$ and the complex angle $\hat \varrho = a+ib$ that enters the matrix $\mathcal{R}$.

The contour plots of Fig.~\ref{fig:yl-scans} show the leptogenesis scale $M^2/\mu$ that yields the experimental baryon asymmetry, $Y_B =8.7\times 10^{-11}$, for given values of $a$ and $b$, and in the NH (a) and IH (b) scenarios. For the NH case, a large region of the parameter space yields successful leptogenesis at the PeV scale. The IH scenario requires a slightly larger scale, $M^2/\mu\gtrsim 6$ PeV. This is because the imaginary part of $YY^\dagger_{12}$ scales with the light neutrino mass difference, which is smaller for IH.

 Eq.\ (\ref{eq:mgluino}) and the experimental bound $m_{\tilde{g}}\gtrsim 2.3$~TeV put a lower bound on the
ratio $F_X/\widetilde{X}$. We also recall that  stability of the $\sn = \sn'=0$ vacuum requires $\lambda' F_X/M^2\approx \lambda' F/M^2 \leq 1$. Combining these constraints, 
one obtains a lower bound on the mass scale of the $\bm{N}'_i$ states,
\be \label{eq:leptoscale}
\frac{M^2}{\mu} \gtrsim 700\ {\rm TeV}\  \lp 0.1\over s_M\rp\ \lp \frac{M^2}{\lambda' F_X} \rp
\ee 
where we used $g_s^2/4\pi \approx 0.12$. Eq.~(\ref{eq:leptoscale}) can be understood as an absolute lower bound for the scale of leptogenesis in our model, which is orders of magnitude below the $\sim10^9$~GeV Davidson-Ibarra bound in standard leptogenesis \cite{Davidson:2002qv} and the $\sim 10^7$~GeV limit seen in soft leptogenesis \cite{Fong:2011yx} and other models with radiatively-induced light neutrino masses \cite{Ma:2006ci}.

\subsection{Charged lepton flavor violation}
SUSY-breaking models typically open the way to charged lepton flavor violating processes which are highly constrained, such as $\mu\to e\gamma$ \cite{Casas:2001sr,Hisano:1995cp,Kashti:2004vj}. These interactions are allowed by off-diagonal entries in the slepton soft mass matrix, 
\be 
(m^2_\slp)_{\alpha\beta}\approx -\frac{3 m_0^2}{8\pi^2}\, (Y^\dagger Y)_{\alpha\beta}\, \ln\lp M_{\rm GUT}\over M \rp,
\ee 
where $m_0$ is the universal slepton mass.
The branching ratio of $L_\alpha\to L_\beta \gamma$ decays is approximately given by
\be 
\mathrm{BR}\left(L_\alpha \to L_\beta \gamma\right) \approx \frac{\alpha^3}{G_F^2} \frac{|(m_\slp^2)_{\alpha \beta}|^2}{m_{\mathrm{SUSY}}^8} \tan ^2 \beta,
\ee 
where $\alpha$ is the fine structure constant, $G_F$ is the Fermi constant and $m_{\mathrm{SUSY}}$ is the superpartner mass scale. 

Assuming $m_0\approx m_{\mathrm{SUSY}}$, we can numerically estimate this branching ratio in our leptogenesis model,
\bea 
\mathrm{BR}\left(L_\alpha \to L_\beta \gamma\right) &\approx& 4.5\times 10^{-19} \lp \frac{(Y^\dagger Y)_{\alpha\beta}}{1.6\times 10^{-6}} \rp^2 \nn\\ 
&&\times \lp \frac{\mathrm{1\ TeV}}{m_{\rm SUSY}} \rp^4  \lp \frac{\tan \beta}{10} \rp^2,
\eea
where we used $M_{\mathrm{GUT}}\approx 10^{16}$~GeV and $M\approx10$~PeV. This estimate is orders of magnitude below the current bound for lepton flavor violation in muon decays, $\mathrm{BR}(\mu\to e\gamma)\lesssim4.2\times10^{-13}$ \cite{MEG:2016leq}, and well below the $\sim 6\times 10^{-14}$ predicted sensitivity for the MEG~II experiment \cite{Meucci:2022qbh,MEGII:2021fah}. Charged lepton flavor violation in our model is therefore too small to be observed in the near future, which is a general consequence of the flavor blindness of gauge-mediated SUSY breaking models \cite{Martin:1997ns,Giudice:1998bp}.

\begin{table*}
\centering
\begin{tabular}{|c | c | c | c | c | c | c | c | c | c | c | c | c ||c|c|c|c|} 
 \hline
Model & $\sqrt{F}$ & $m$ & $\lambda$ & $g$ & $\sqrt{D}$ & $T_n$ & $\langle x\rangle_{T=T_n}$ & $\alpha$ & $\beta_H$ & $\sqrt{F_X}$ & $m_{\tilde g}$  (TeV) & $\epsilon_R$  & $\lambda'$ & $M$ & $\mu$ & $M^2/\mu$  \\
\hline
1 &30&  50.7 & 2.14 & 0.1 & 84.0 & 26.9 & 769 & 0.429 & 76.8 & 24.6 & $2.49$ & $1.90\times10^{-3}$ & 1.58& 40 & 890& 1.80  \\
\hline
2 & 30& 59.7 & 2.97 & 0.117 & 78.3 & 26.9 & 555 & 0.317 & 74.4 & 21.5 & $2.58$ & $2.69\times10^{-3}$&0.623 &40& 258 &6.20 \\
\hline
\end{tabular}
\caption{Parameters for two benchmark models with successful leptogenesis and high GW production. All the dimensionful quantities are expressed in PeV except for the gluino mass. }
\label{tab2}
\end{table*}

\section{Other cosmological signatures}
\label{sec:signatures}
In addition to providing a mechanism for leptogenesis, a first-order SUSY-breaking phase transition may produce other cosmological signals that future experiments can observe. In this section, we start by estimating the GW spectrum generated by such a FOPT and compare it to the expected noise spectrum of several proposed GW detectors. We then demonstrate that this SUSY-breaking phase transition would be prone to producing primordial black holes.

\subsection{Gravitational waves}
\begin{figure*}[t]
    \centering
    \includegraphics[width=0.5\linewidth]{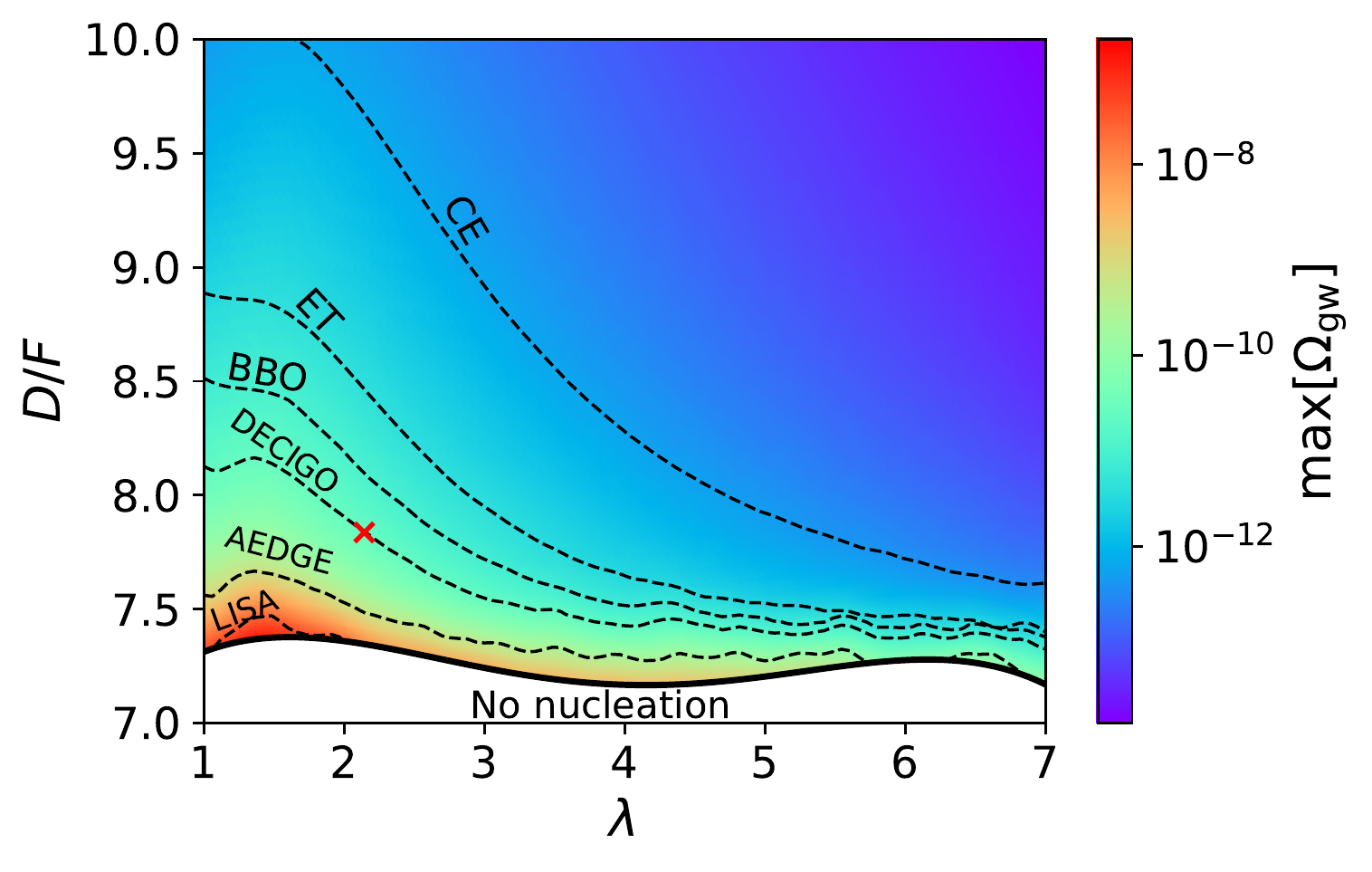}\includegraphics[width=0.5\linewidth]{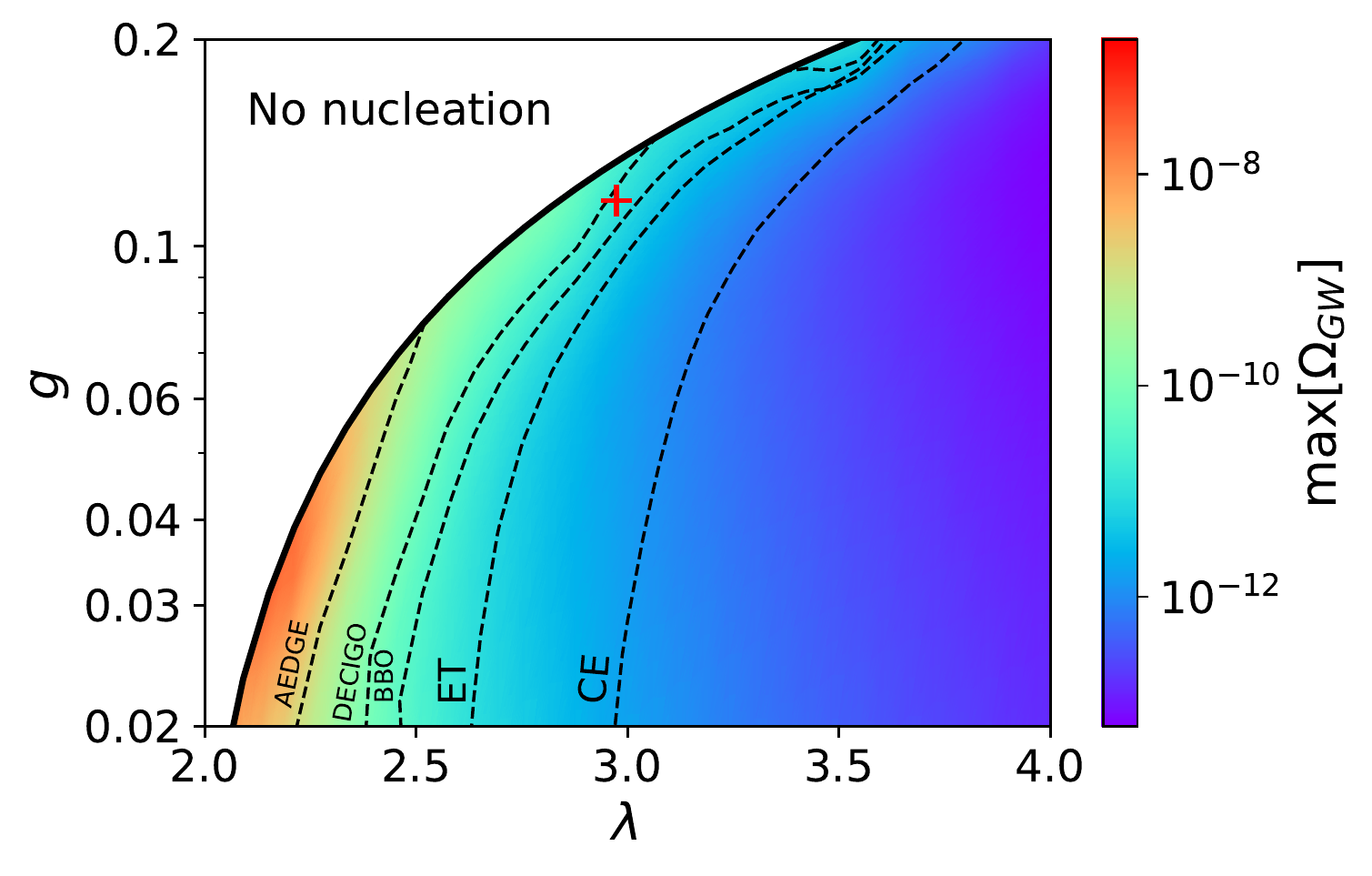}\\[-3mm]
    \textbf{(a)}\hspace{0.48\linewidth}\textbf{(b)}\hspace{0.35\linewidth}
    \caption{Intensity plot of maximal amplitude of the GW spectrum. The regions (a) below or (b) left of the dashed lines would be detected by the indicated experiment. We fixed (a) $\sqrt{F}=30\ \mathrm{PeV}$, $g=0.1$ and $\lambda F/m^2=3/4$, and (b) $\sqrt{F}=30\ \mathrm{PeV}$, $\lambda F/m^2=3/4$ and $gD/m^2=1/5$. The red signs $\times$ and $+$ show respectively the position of the models 1 and 2 of Table \ref{tab2}.}
    \label{fig:scans}
\end{figure*}

To assess the observability of GWs produced in the FOPT, one
considers the  spectrum $\Omega_{\gw}(f)$, which is the contribution per frequency octave to the energy density in gravitational waves, \textit{i.e.}, $\int\Omega_{\gw}(f)\ d(\log f)$, the fraction of energy density compared to the critical density of the universe. In general, the spectrum can be separated into contributions from the scalar fields, sound waves in the plasma and magnetohydrodynamical turbulence. However, the scalar field contribution is only important for runaway walls ($\gamma_w\rightarrow\infty$); Ref.\ \cite{Bodeker:2017cim} showed that for ultrarelativistic walls, interactions with gauge bosons create a pressure on the wall proportional to $\gamma_w$ which prevents it from running away. Furthermore, the estimates for the magnetohydrodynamical turbulence are  uncertain and sensitive to the details of the phase transition dynamics \cite{RoperPol:2019wvy}, and are expected to be much smaller than the contribution from sound waves. Hence, we only consider the contribution from the latter.

The GW spectrum from sound waves observed today can be parameterized as \cite{Hindmarsh:2020hop}
\be\label{eq:GWspectrum}
\Omega_{\gw}(f) = 2.061 F_{\gw,0}\tilde\Omega_\gw\frac{(HR)^2}{\sqrt{K}+HR}\,K^2\,C(f/f_{\rm p,0})\,,
\ee
where $F_{\gw,0}=3.57\times10^{-5}\lp{100}/{g_*}\rp^{1/3}$, quantifies the decrease in GW energy from the expansion of the universe,
$R = (8\pi)^{1/3}v_w/{\beta}$ is the mean bubble radius at the time of percolation,\footnote{We make the approximation $T_*\simeq T_n$, where $T_*$ is the percolation temperature.}
$K=\kappa\alpha/({1+\alpha})$ is the kinetic energy fraction,
$C(s)= s^3\lp{7}/({4+3s^2})\rp^{7/2}$ is a function determined from simulations that approximate the spectrum's shape,
and the peak frequency is
\be
f_{\rm p,0} = 2.62\lp\frac{1}{HR}\rp\lp\frac{T_n}{100\ \mathrm{PeV}}\rp\lp\frac{g_*}{100}\rp^{1/6}\ \mathrm{Hz}\,.
\ee
Furthermore, $g_*$ is the effective number of degrees of freedom, $\beta$ and $\alpha$ were given in Eqs.\ (\ref{eq:beta}-\ref{eq:alpha}), $v_w$ is the wall velocity,\footnote{Determining $v_w$ is a notoriously difficult problem; hence we do not try to do a complete calculation here. Nevertheless, Refs.\ \cite{Cline:2021iff,Laurent:2022jrs} showed that for strong FOPTs ($\alpha\gtrsim 0.01$), the wall  becomes ultrarelativistic. Since our model always yields $\alpha\gtrsim0.1$,  $v_w\cong1$ and we adopt this value.} $\tilde\Omega_\gw=0.012$ is a constant determined numerically which represents the efficiency with which kinetic energy is converted into GWs, and $\kappa\cong {\alpha}/({0.73+0.083\sqrt{\alpha}+\alpha})$ is the efficiency with which vacuum energy is turned into kinetic energy.

Once the GW spectrum is known, it must be compared to the sensitivity of a detector to assess its detectability. The signal-to-noise ratio (SNR) is defined by
\be
\SNR = \sqrt{\mathcal{T}\int_{f_{\rm min}}^{f_{\rm max}}\!df\lc\frac{\Omega_\gw(f)}{\Omega_{\rm sens}(f)}\rc^2},
\ee
where $\Omega_{\rm sens}(f)$ denotes the sensitivity curve of the detector and $\mathcal{T}$ is the duration of the mission. Whenever SNR is greater than a given threshold $\SNR_{\rm thr}$, we conclude that the signal can be detected. In general, the threshold depends upon the configuration of the detector and can be complicated to compute, but the value $\SNR_{\rm thr}=10$ is reasonable for most detectors.
We compare the prediction (\ref{eq:GWspectrum}) for a range of parameters to several proposed GW detectors: the earth-based detectors LIGO \cite{LIGOScientific:2014pky}, ET \cite{Sathyaprakash:2012jk} and CE \cite{LIGOScientific:2016wof}, and the space-based detectors LISA \cite{Audley:2017drz,Robson:2018ifk}, AEDGE \cite{AEDGE:2019nxb}, DECIGO \cite{Seto:2001qf,Yagi:2013du} and BBO \cite{Crowder:2005nr,Yagi:2013du}. For each experiment, we assume $\SNR_{\rm thr}=10$ and $\mathcal{T}=4\ \mathrm{years}$, and the sensitivity curves can be found in the previous references.

We performed two scans of the parameter space at a scale $\sqrt{F}=30\ \mathrm{PeV}$ and computed the SNR for each detector. In both scans, $\lambda$ ranges from 1 to 7, with $D/F\in[6,13]$ in the first scan and $g\in[0.02,0.2]$
in the second.
Fig.\ \ref{fig:scans} shows the amplitude of the GW signal in the planes of the varied parameters, and the regions of sensitivity of the future experiments.
It demonstrates that a large region of parameter space can be probed by the proposed experiments, especially by the earth-based detectors ET and CE. In contrast, no model is detectable by LIGO. Fig.\ \ref{fig:scans}  shows that the GW amplitude is maximized at small couplings $\lambda\lesssim 2$, $g\lesssim 0.1$ and for $D/F\lesssim 8$. The amplitude is maximal at the boundary of the `No nucleation' region, where no solution to Eq.\ (\ref{eq:Tn}) exists. Close to this boundary, the phase transitions have enhanced supercooling, which leads to a stronger FOPT and consequently a larger GW amplitude.

\begin{figure}[t]
    \centering
    \includegraphics[width=1\linewidth]{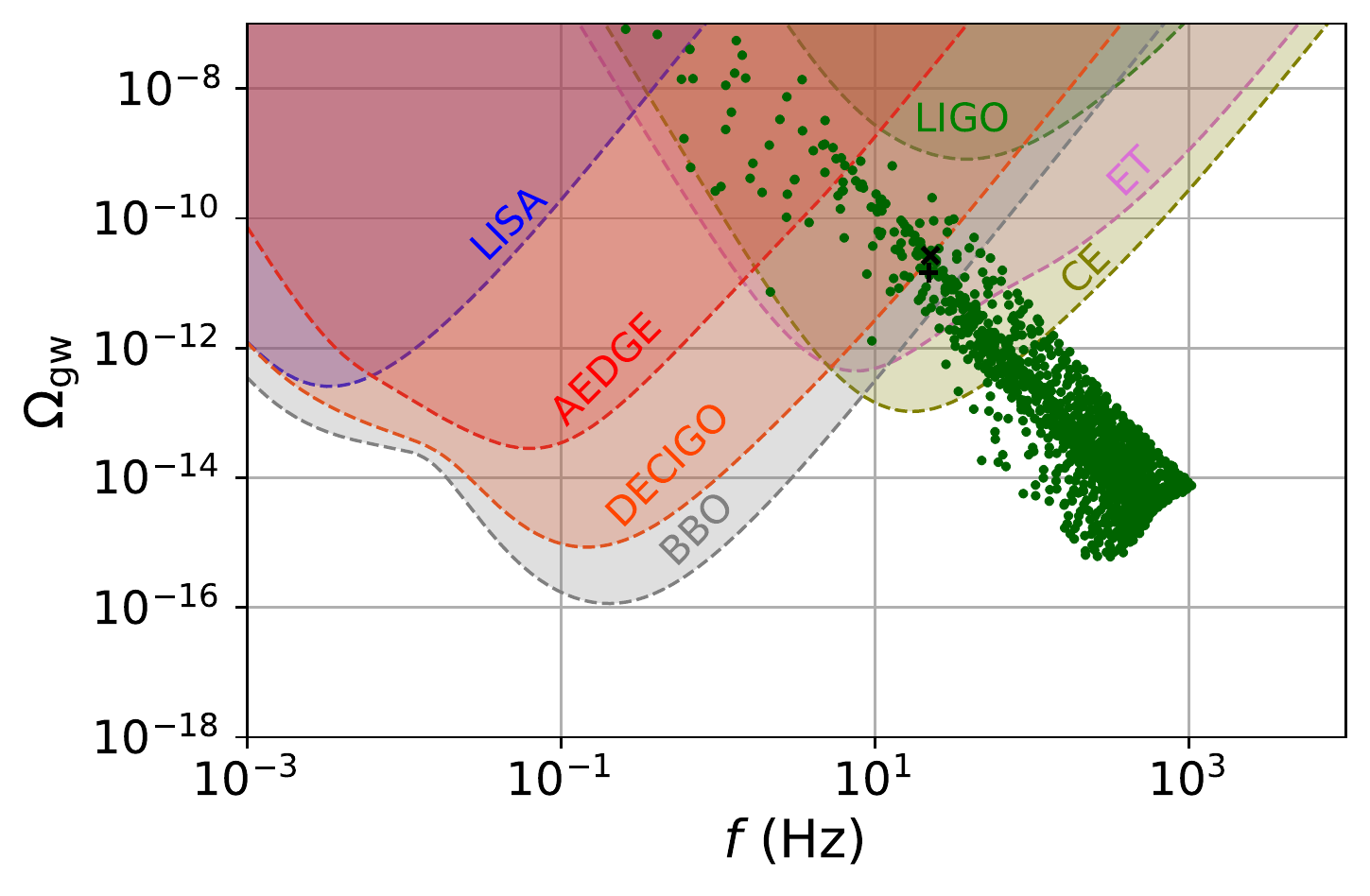}
    \caption{Peak-integrated sensitivity curves of the detectors (background) and peak values of the GW signal (foreground) for the models used to generate Fig.\ \ref{fig:scans}. The signs $\times$ and $+$ show respectively the position of the models 1 and 2 of Table \ref{tab2}.}
    \label{fig:pisc}
\end{figure}

Fig.\ \ref{fig:pisc} shows the peak-integrated sensitivity curves \cite{Schmitz:2020syl} (PISC) of the detectors,
along with the peak value of the GW spectrum for each  model appearing in Fig.\ \ref{fig:scans}. The PISC is defined in such a way that any GW signal whose amplitude peaks above the PISC will be detected. It is therefore an intuitive figure of merit for the sensitivity of a detector. 
At the scale $\sqrt{F}=30\ \mathrm{PeV}$, the frequency range of the GW signal produced by the FOPT coincides with the region of peak sensitivity of the earth-based detectors. This allows them to probe a larger region of parameter space, despite their lower sensitivity relative to space-based detectors. If we lower the scale to $\sqrt{F}=3\ \mathrm{PeV}$, the peak frequency is also rescaled by the same factor since all the dimensionful parameters are expressed as a ratio of $\sqrt{F}$\footnote{Only the Hubble parameter $H$, which appears in Eq.\ (\ref{eq:nucleation}), has a different scaling relation $H\propto F/m_P$. However, $T_n$ only depends logarithmically on $H$, so it has a small effect and one still approximately has $T_n\propto\sqrt{F}$.}, so we must have $f_{\rm p,0}\propto \sqrt{F}$. The frequency range then becomes closer to the sensitivity region of the space-based detectors. This highlights the importance of having both types of detectors to cover a large range of frequencies, and of the corresponding energy scales.

\subsection{Primordial black holes}

\begin{figure*}[ht]
    \centering
    \includegraphics[width=0.5\linewidth]{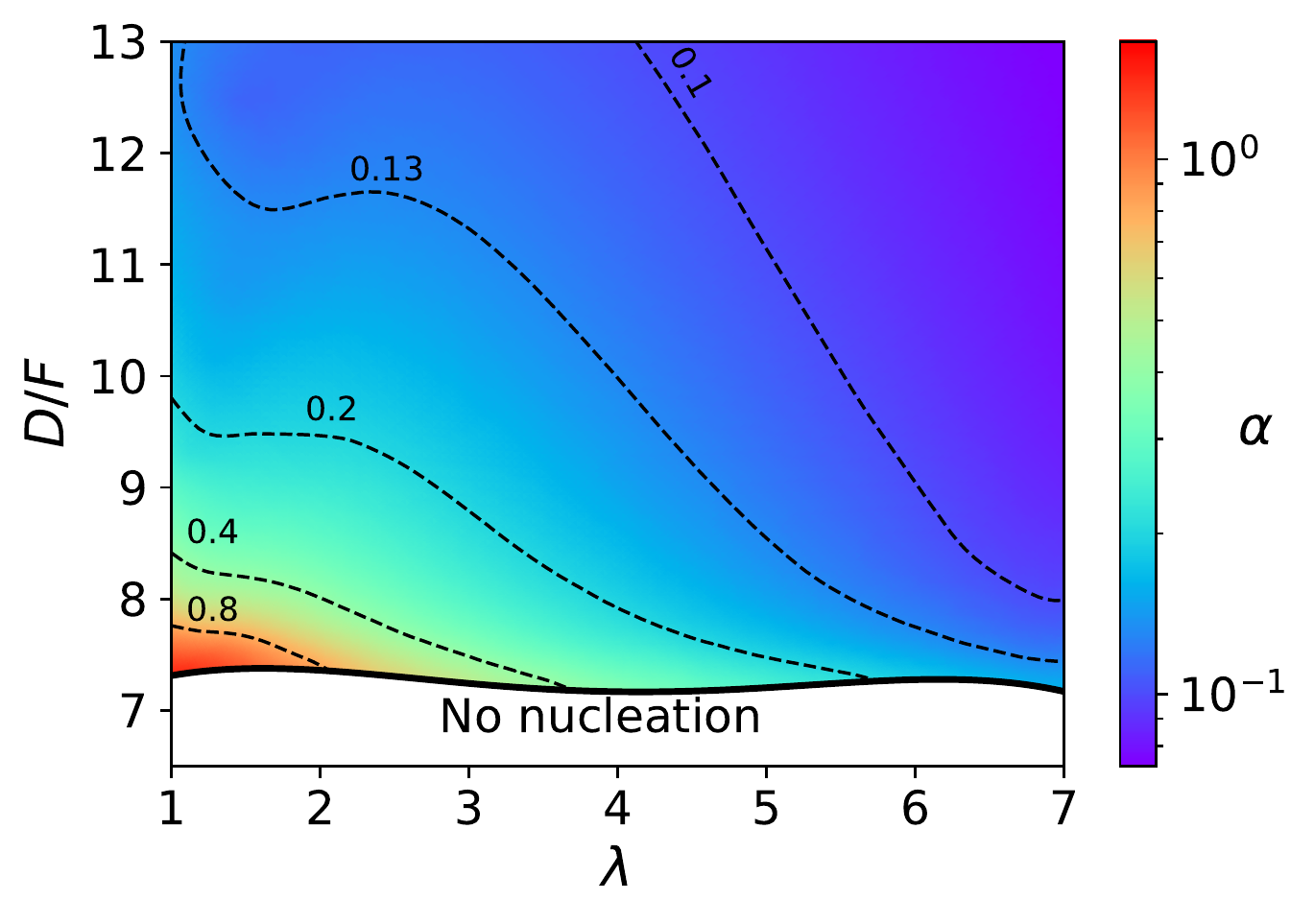}\includegraphics[width=0.5\linewidth]{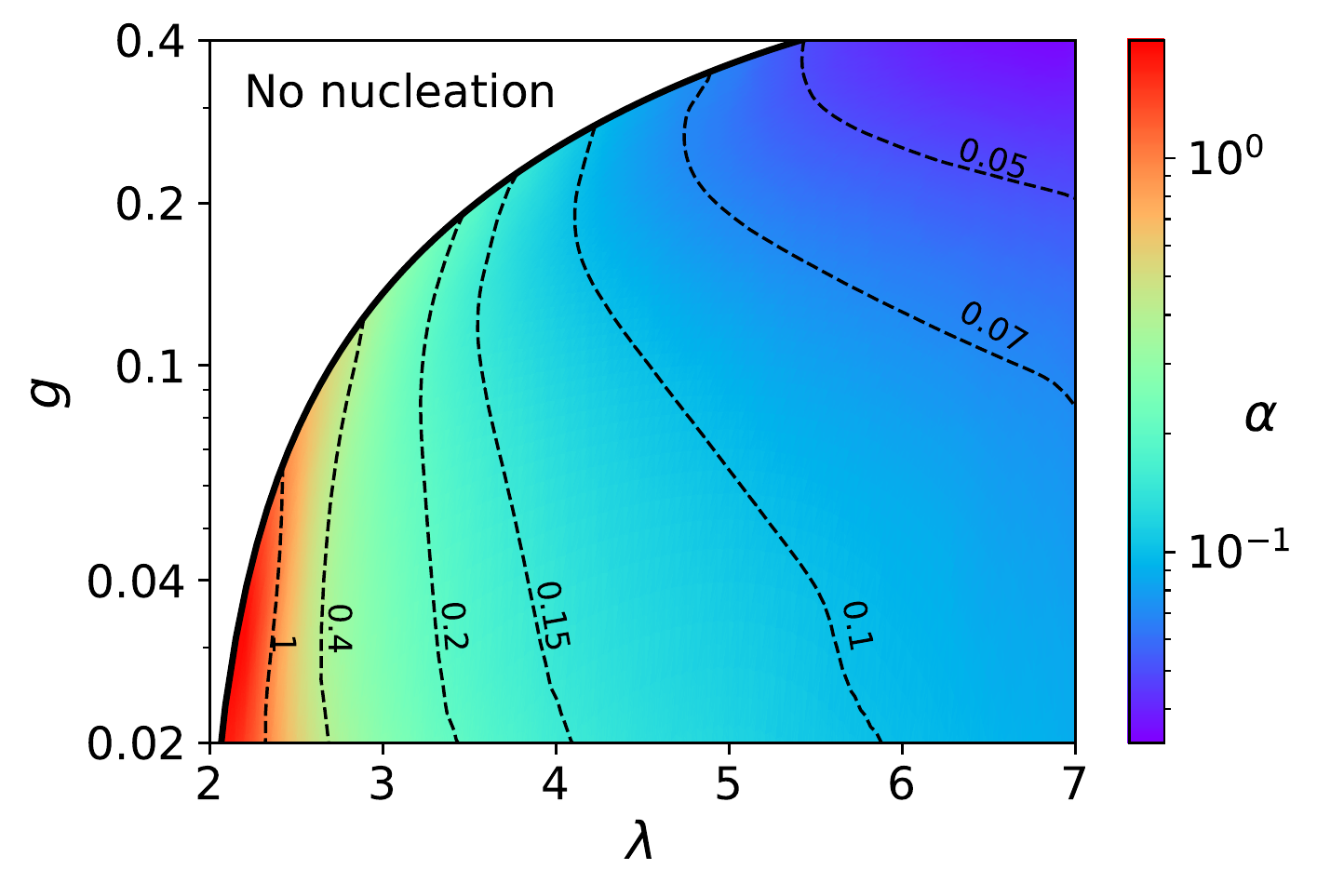}\\
    \includegraphics[width=0.5\linewidth]{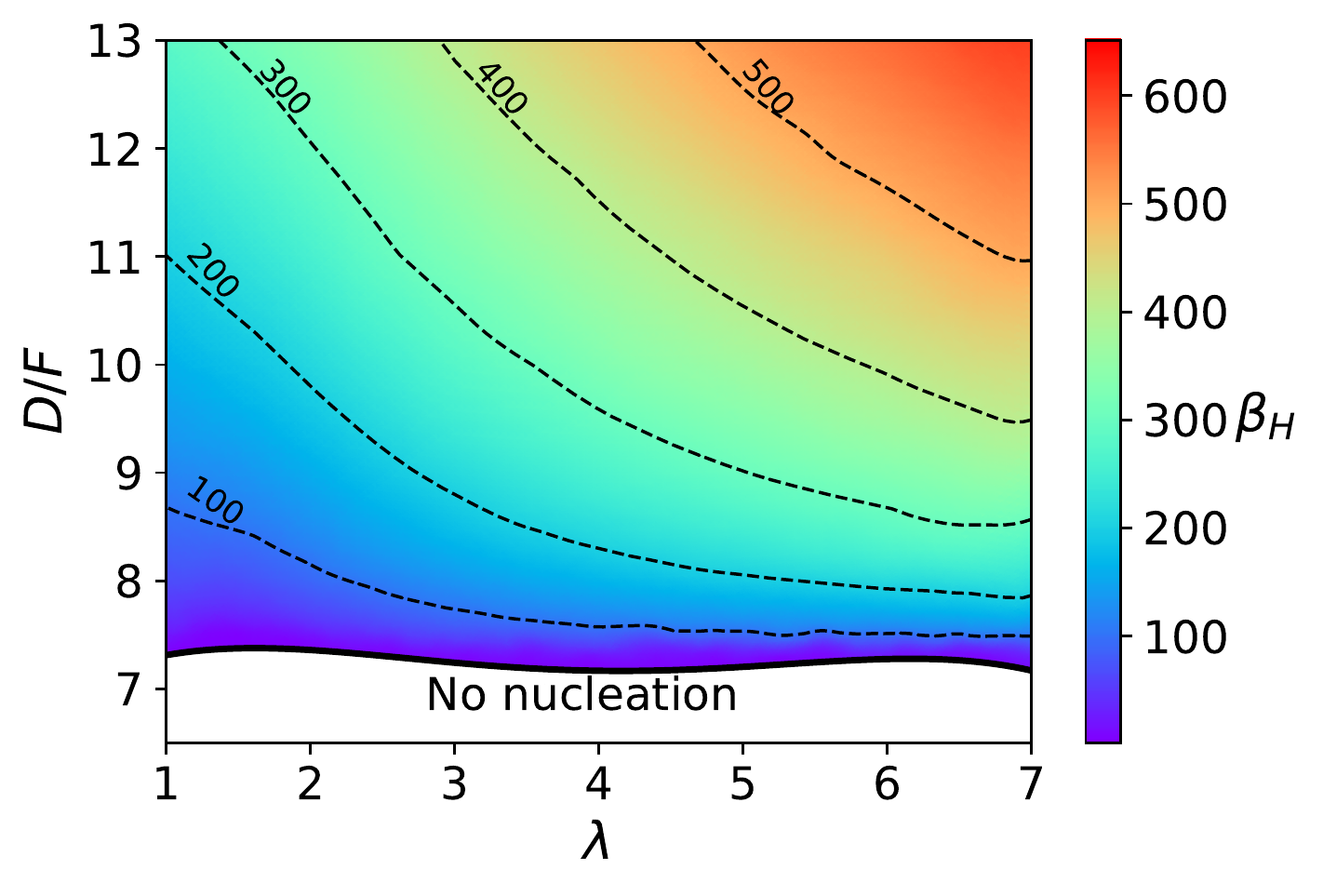}\includegraphics[width=0.5\linewidth]{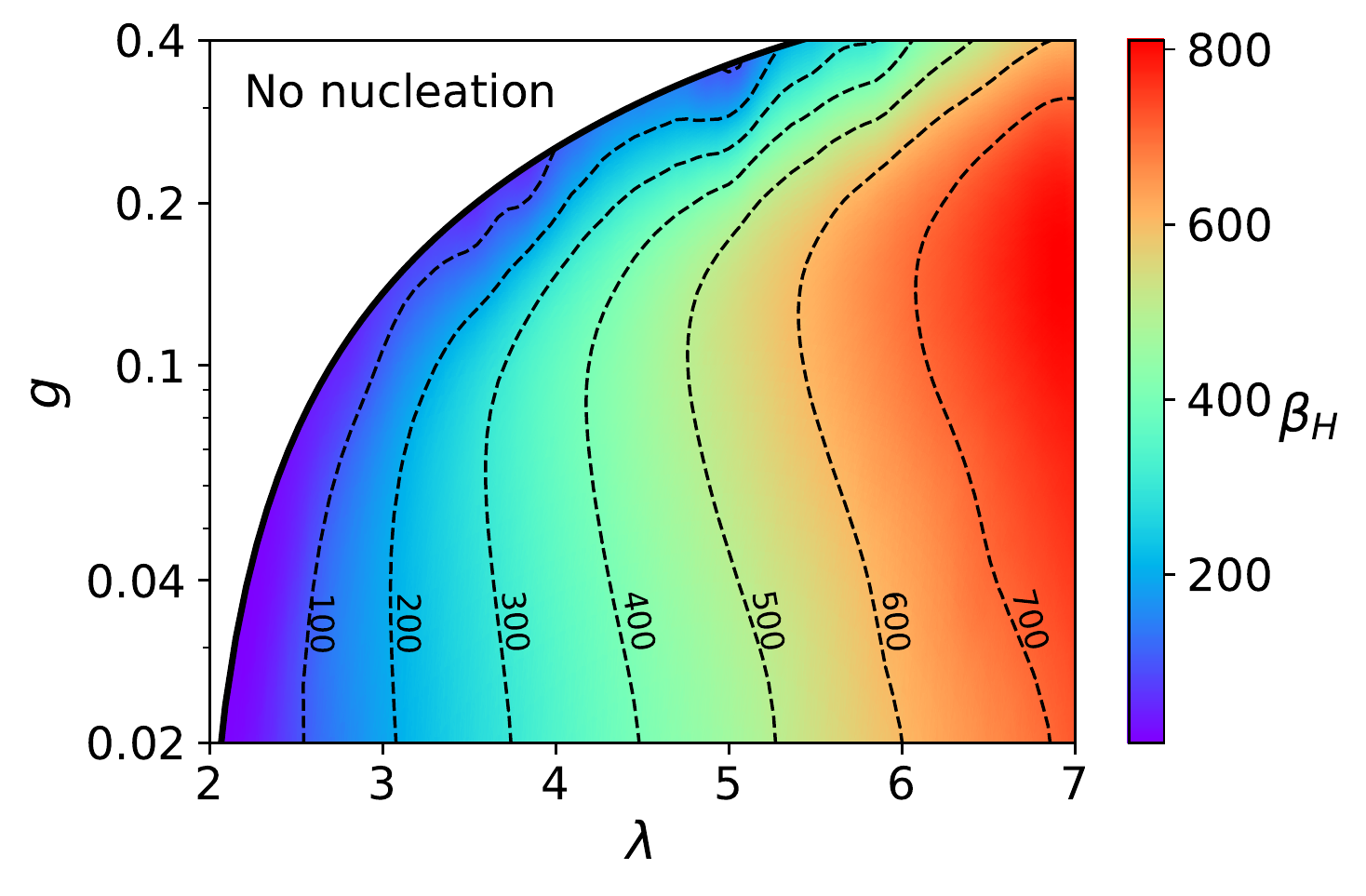}\\[-3mm]
    \textbf{(a)}\hspace{0.48\linewidth}\textbf{(b)}\hspace{0.35\linewidth}
    \caption{$\alpha$ and $\beta_H$ intensity plots for the scans presented in Fig.\ \ref{fig:scans}}
    \label{fig:alpha-beta}
\end{figure*}

Several recent studies suggest that primordial black holes (PBHs) can be produced by collision of true-vacuum bubbles during a FOPT. One mechanism relies on the energy stored in the wall to cause gravitational collapse when neighboring bubbles collide \cite{Jung:2021mku}. However, to reach the required energy density, the bubbles must have a large radius, which implies $\beta_H\ll1$. In our model, obtaining such values of $\beta_H$ requires significant fine tuning, since typically $\beta_H\sim100$, and the smallest value from the 2000 models represented in Fig.\ \ref{fig:scans} was $\beta_H\sim1$.

A more promising mechanism in the present context is mass gain by particles across the bubble walls \cite{Baker:2021nyl,Baker:2021sno,Huang:2022him}. It requires a species that is initially light to acquire a large mass $\Delta m\gg \gamma_w T$ during the FOPT. These particles do not have enough kinetic energy to go through the wall and are trapped in the false vacuum. At the end of the phase transition, they form false-vacuum bubbles that are compressed by the vacuum pressure and can thereby lead to a gravitational collapse.

As argued in Section \ref{sec:FOPT}, the pseudomodulus field $x$ naturally gets a large VEV during the phase transition; hence it is generic to have large variation of masses across the wall. For example, if $m\lesssim T$ and $\lambda \langle x\rangle\gg\gamma_w T$, half of the messenger fields\footnote{For simplicity, we only consider the contribution from the messengers, but the same mechanism would work with the neutrinos $N_i$ if $M\lesssim T$ and $\lambda' \langle x\rangle\gg\gamma_w T$.} would gain a large enough mass to contribute to the collapse (the other half becoming light and not trapped in the false-vacuum bubble). 

Determining the final abundance and mass spectrum of the resulting PBHs is a complicated task that is beyond the scope of this paper. Nevertheless, one can follow the methodology of Ref.\ \cite{Baker:2021sno} to estimate that PBH production should be efficient. For simplicity, we take the false-vacuum bubble at the end of the phase transition to be spherical, with radius $r(t)$. One can show that, for nonrelativistic walls,\footnote{We use the nonrelativistic limit to simplify the analysis. In reality, the wall is ultrarelativistic and therefore transmits more energy to the particles that reflect on it. Thus, the estimates derived here are conservative and one should expect the actual PBH production to be larger.}, the energy density of the heavy messengers inside the bubble scales like $r(t)^{-4}$, assuming that none of them can escape the bubble.

For the bubble to collapse into a black hole, $r(t)$ must become smaller than the Schwarzschild radius $r_s$:
\be
r(t) < r_s = 2GE_{\rm tot},
\ee
where $E_{\rm tot}$ is the total energy of the heavy messengers inside the bubble. Using the previous scaling relation for the energy density, one can show that this condition leads to 
\be
\frac{r(t)}{r_H} < \sqrt{\frac{a g_\Psi}{g_*}}\lp\frac{r_0}{r_H}\rp^2,
\ee
where $g_\Psi=45$ ($g_*\cong341.25$) is the heavy messenger (total) effective number of degrees of freedom, $r_0=r(t_0)$ is the initial bubble radius and $r_H=1/H$ is the Hubble radius. 

We have also introduced the parameter 
\be
a\equiv \frac{\rho_\Psi(r_0)}{\rho_{n}}\,,
\ee
where $\rho_{n}=\pi^2g_\Psi T_n^4/30$ is the messenger's thermal energy density at the beginning of the phase transition and $\rho_\Psi(r_0)$ is the energy density when the false-vacuum bubbles form. In general, one expects $a>1$ since when the bubbles form, there is already a nonnegligible fraction of the universe in the true vacuum, which decreases the volume of the false vacuum regions, and thereby increases their energy density. Assuming that the energy scales like $V^{-4/3}$ and that the first false vacuum bubbles form at the time of percolation, when the volume fraction remaining in the false vacuum is $p_f\cong 0.71$ \cite{Rintoul_1997}, we estimate $a\cong p_f^{-4/3}\cong 1.6$. For a typical initial size of $r_0=r_H$, the bubble need only shrink by a factor of 2.2 to collapse.

One must still assess whether the bubble can shrink this much. A necessary condition is that the net inward pressure remains positive while $r(t)>r_s$. Assuming that the plasma remains in thermal equilibrium and that the bubble shrinks adiabatically without losing particles, the thermal pressure which opposes the contraction is given by
\be
P_T(r) = \frac{1}{3}\rho_\Psi(r) = \frac{a\rho_{n}}{3}\lp\frac{r_0}{r(t)}\rp^4.
\ee
This must be compared to the 
vacuum pressure $\Delta V$ and the gravitational pressure, 
both of which promote the collapse.  
For a uniform spherical bubble, the gravitational energy is
\bea
E_G&=&-\frac{3GE_\Psi^2}{5r} 
= -\frac{3G}{5r}\lp\frac{4}{3}\pi r^3\rp^2(a\rho_n)^2\lp\frac{r_0}{r}\rp^8\nn\\
&=&-\frac{2\pi}{5r^3}r_0^8 H^2 a^2\rho_n\lp\frac{g_\Psi}{g_*}\rp \,,
\eea
yielding the gravitational pressure
\be
P_G=\frac{\partial E_G}{\partial V}=\frac{3r_0^8}{10r_H^2r^6}\,a^2\rho_n\lp\frac{g_\Psi}{g_*}\rp \,.
\ee

It follows that the net pressure $P_{\rm net}=\Delta V+P_G-P_T$ is minimized for a false vacuum bubble of radius
\be
r_{\rm min} = \frac{3}{2}\sqrt{\frac{3a g_\Psi}{5g_*}}\frac{r_0^2}{r_H}\cong 1.162\, r_s\,.
\ee
Hence, the most stringent constraint does not arise from $r(t)$ reaching the Schwarzschild radius $r_s$, but rather at the slightly larger radius $r_{\rm min}$. Once a bubble becomes smaller than $r_{\rm min}$, the gravitational pressure starts to dominate and collapse is inevitable. The condition on the net inward pressure at $r_{\rm min}$ for collapse is
\be
P_{\rm net}(r_{\rm min}) = \Delta V-\frac{400}{6561\,a}\lp\frac{g_*}{g_\Psi}\rp^2\lp\frac{r_H}{r_0}\rp^4\rho_n>0\,.
\ee
Neglecting the second term in the definition (\ref{eq:alpha}) of $\alpha$, one then obtains the condition
\be\label{eq:PBHBound}
\alpha > \frac{400}{6561a}\lp\frac{g_*}{g_\Psi}\rp\lp\frac{r_H}{r_0}\rp^4 \cong 0.29 \lp\frac{r_H}{r_0}\rp^4
\ee
for the formation of PBHs.

The criterion (\ref{eq:PBHBound}) implies that only very strong FOPTs can produce PBHs. For weaker transitions, the thermal pressure grows too rapidly as the false-vacuum bubble shrinks, and it eventually overcomes the vacuum and gravitational pressures. It is also apparent that PBH production is favored in large initial bubbles. Since one expects $r_0/r_H\sim 1/\beta_H$, $\beta_H$ should not be too large. Intensity plots of $\alpha$ and $\beta_H$ are shown in Fig.\ \ref{fig:alpha-beta}, that demonstrate
the existence of extended regions with large $\alpha$, close to the
no-nucleation boundary. These correspond to the regions of minimal $\beta_H$, which favors PBH production. 

We emphasize that the heuristic analysis made here neglects several physical effects. For example, for relativistic walls, the particles gain more energy with each collision, making it easier to satisfy the Schwarzschild radius criterion. On the other hand, several processes can reduce the number of particles in the bubbles (\textit{e.g.,} decays into light messengers, annihilation into $XX$ pairs, sufficiently energetic particles able to cross the wall), which reduce the energy density. Such effects were investigated in Refs.\ \cite{Baker:2021nyl,Baker:2021sno}. Furthermore, we made the approximation of relativistic messengers, whereas they typically have a mass of $m/T_n\sim 1 - 2$. Recomputing the criterion (\ref{eq:PBHBound}) numerically with a finite mass, we find that it scales approximately as $n^{-5/4}$, where $n\sim g_\Psi T_n^3 e^{-m/T_n}$ is the messenger number density. Therefore, PBH production is strongly Boltzmann suppressed at high $m$, but the bound on $\alpha$ remains reasonably low for $m/T_n\sim 1-2$.  A quantitative calculation of the final PBH abundance would require determining the distribution of the $r_0$ and $a$ value, which is beyond the scope of this paper.

Nevertheless, the criterion (\ref{eq:PBHBound}) is very general. As long as a few fundamental conditions are satisfied---namely $\Delta m\gg \gamma_w T$, $v_w\ll 1$ and that the number of particles stays roughly constant---efficient PBH production 
should be predicted by a condition similar to (\ref{eq:PBHBound}), irrespective of the details of the phase transition.

\section{Conclusions}
\label{sec:conclusion}
In this work we have developed the paradigm started in Ref.\ \cite{Craig:2020jfv}, where the potential for first-order supersymmetry breaking phase transitions to produce observable gravity waves was initially explored.  A primary challenge undertaken here was to extend the original framework to encompass
viable leptogenesis, to simultaneously explain the baryon asymmetry of the universe.  This proved to be more constraining than might be expected {\it a priori}, due to our hypothesis
that the asymmetry could be linked to the phase transition.

In particular, we assumed that lepton number is broken by a single interaction $\lambda' XN_iN_i$ coupling heavy sterile neutrinos to the pseudomodulus field in the superpotential $W$,\footnote{ The corresponding bare mass term $MNN$
is forbidden in $W$ by $R$ symmetry.} which leads to the heavy neutrino mass $m_N=\lambda'\langle\tilde X\rangle$ being correlated with the scale of SUSY breaking, taken to be $\sim 10\,$PeV to get observable gravity waves.  This is too low for
conventional leptogenesis.  We found these challenges could be overcome by introducing a second set of heavy neutrinos $N'_i$ that pair with $N_i$ to form Dirac states before SUSY breaking (but become lighter than $N_i$ after SUSY breaking), and whose out-of-equilibrium decays, along with those of the corresponding sneutrinos, produce the lepton asymmetry.  In this setup, light neutrino masses vanish at tree level, but get generated at one loop via virtual $Z$-$N$ exchange.  A large enough lepton asymmetry is achieved by assuming minimal flavor violation in the leptonic sector, which makes the
heavy $N'_i$ nearly degenerate, leading to partially resonant
leptogenesis.  This occurs at the scale $T\lesssim\sqrt{F}$, which is much lower than in conventional leptogenesis.

A novel outcome is our proposal for neutrino mass generation at one loop, due to the presence of additional right-handed neutrino species $N_i'$ that cause the tree-level
masses to vanish.  The resulting neutrino mass spectrum
is similar to that provided by the seesaw mechanism, 
with an effective right-handed neutrino mass that is parametrically larger than the actual mass by a factor of 
$2\pi/\alpha_w$.

Ref.\ \cite{Craig:2020jfv} noted that the SUSY-breaking scale $\sqrt{F}$ is rather narrowly constrained, since LHC limits on the gluino mass bound it from below, while gravitino overproduction, combined with Big Bang Nucleosythesis, bounds it from above.  The LHC constraint is strengthened in our model, which predicts a definite degree of gaugino screening, pushing the gluino mass close toward its current limit.  In this sense the model is quite predictive, requiring $m_{\tilde g}$ to be not much higher than 2 TeV.

We have taken a preliminary step toward estimation of the primordial black holes, by the mechanism of particles being trapped in the disappearing false vacuum regions toward the end of the phase transition.  A full study of the spectrum of produced PBHs would require simulating the phase transition on a lattice, which is beyond the scope of the present work.   The criterion we derived for which false vacuum bubbles would lead to PBH formation may be useful in such a future investigation.

\bigskip
{\bf Acknowledgements.}  This work was supported by the Natural Sciences and Engineering Research Council (NSERC) of Canada and the Fonds de recherche du Qu\'ebec Nature et technologies (FRQNT).
We thank M.\ Baker, N.\ Craig, G.\ Giudice,  T.\ Han,  E.\ Ma, R.\ Mohapatra  and D.\ Redigolo for helpful correspondence or feedback.

\begin{appendix}
\section{One-loop potential and mass eigenvalues}
\label{appA}

One-loop corrections are given by the Coleman-Weinberg potential:
\be \label{eq:CWpot}
V_{CW}(\phi_i) = \sum_j (-1)^F \frac{g_j m_j^4(\phi_i)}{64\pi^2} \left(\log(\frac{m_j^2(\phi_i)}{Q^2})-c_j\right),
\ee 
where $F=0$ ($1$) for bosons (fermions), $g_i=1$ (2) for scalars (fermions), $c_i$ = 3/2 for scalars and fermions in the $\overline{\rm MS}$ renormalization scheme, and $Q$ is the renormalization scale. We will take $Q=m$. Here $\phi_i$ indicates the dependence of the mass eigenvalues on $x$ and the VEVs of the $U(1)_D$ messenger fields. The sum is taken over all tree-level mass eigenvalues in our model.

Similarly, the thermal corrections to the potential are given by
\be\label{eq:thermalpot}
V_T(\phi_i,T) = \frac{T^4}{2\pi^2}\sum_j(-1)^F g_j J_{B/F}\lp\frac{m_j^2(\phi_i)}{T^2}\rp,
\ee
with the thermal functions
\be
J_{B/F}(z^2)=\int_0^\infty dx x^2\log\lc 1-(-1)^F\exp(-\sqrt{x^2+z^2}) \rc.
\ee
At high temperature, these functions can be approximated to lowest order by the following expansions:
\bea\label{eq:highT}
J_B(z^2) &=&-\frac{\pi^4}{45}+\frac{\pi^2}{12}z^2+\cdots\nn\\
J_F(z^2) &=&\frac{7\pi^4}{360}-\frac{\pi^2}{24}z^2+\cdots
\eea

To find the  mass eigenvalues, we write the quadratic terms of the potential as
\be \nn
\frac{1}{2} \begin{pmatrix}
	\phi^* & \phi
\end{pmatrix} \mathbf{m_s^2}(\phi_i) \begin{pmatrix}
\phi \\ \phi^*
\end{pmatrix}
\ee 
and diagonalize the $x$-dependent matrix $\mathbf{m_s^2}(\phi_i)$.

To get the fermionic mass eigenvalues, we must diagonalize the matrix
\be \nn
(\mathbf{m_f})_{ij} =  \left. \frac{\delta^2 W}{\delta \Phi_i \delta\Phi_j}\right|_{\Phi_i\to\phi_i}.
\ee 
The notation $\Phi_i\to\phi_i$ means we replace the superfields with their scalar components.

At tree-level, the components of $X$ and of all MSSM fields are massless, with the exception of the Higgs doublets $H_u,H_d$ that have a mass $\mu$ prior to the breaking of SUSY. However, in the limit where $\mu$ is much smaller than $m$ and $\sqrt{F}$, we can treat the Higgs doublets as massless.

\textbf{Right-handed neutrinos.} The sneutrino mass eigenvalues are given by:
\begin{align}\label{msnp}
\lp m_{\sn_i,\pm}\rp^2 &= M_i^2 + \frac{1}{2}\lp \mu^2 + \Delta_{i,\pm}^2 \pm \lambda' F_X \rp,\nn \\
\lp m_{\sn'_i,\pm}\rp^2 &= M_i^2 + \frac{1}{2}\lp \mu^2 - \Delta_{i,\pm}^2 \pm \lambda' F_X \rp, 
\end{align}
where $\mu = \lambda' x/\sqrt{2}$, $F_X = |F-\lambda \phi_2 \widetilde \phi_2|$ and
\be 
\Delta_{i,\pm}^2 = \sqrt{\mu^4 +4 \mu^2 M_i^2 \pm 2 \mu^2 (\lambda' F_X) + (\lambda' F_X)^2}.
\ee 

The neutrino masses are given by
\begin{align}
m_{N_i} &= \frac{1}{2} \lp \mu + \sqrt{\mu^2 + 4 M_i^2} \rp,\nn\\ 
m_{N'_i} &= \frac{1}{2} \lp \mu - \sqrt{\mu^2 + 4 M_i^2} \rp.
\end{align}

In the limit $\mu\gg M_i$ the superfields $N_i$ and $N'_i$ unmix and split into heavy and light eigenstates, which are approximately given by
\begin{align}
\lp m_{\sn_i,\pm}\rp^2 &\cong \mu^2  +2 M_i^2 \pm \lambda' F_X\nn \\
\lp m_{\sn'_i,\pm}\rp^2&\cong M_i^2\lp\frac{M_i^2 \pm \lambda' F_X}{\mu^2}\rp
\end{align}
for the scalars and
\begin{align}
m_{N_i} &\cong\mu \lp 1+ {M_i^2\over \mu^2}\rp \cong \mu\nn\\
m_{N'_i} &\cong -{M_i^2\over \mu}. 
\end{align}
for the fermions.

\bm{$SU(5)$} \textbf{gauge mediators.} 
The mass matrix of the $SU(5)$ gauge mediators is very similar to that of the right-handed neutrinos. The scalar eigenvalues are
\begin{align} \label{eq:messengermassb}
\lp m_{\widetilde{5}_M,\pm}\rp^2 &= m^2 + \frac{1}{2}\lbr (\lambda x)^2/2 + \Delta_{5,\pm}^2 \pm \lambda F_X \rbr, \nn\\
\lp m_{\widetilde{\overline{5}}_M,\pm}\rp^2 &= m^2 + \frac{1}{2}\lbr (\lambda x)^2/2 - \Delta_{5,\pm}^2 \pm \lambda F_X \rbr,
\end{align}
where 
\be 
\Delta_{5,\pm}^2 = \sqrt{(\lambda x)^4/4 +2 (\lambda x)^2 m^2 \pm (\lambda x)^2 (\lambda F_X) + (\lambda F_X)^2}.
\ee 
These are identical to the sneutrino eigenvalues with $\lambda'\to \lambda$ and $M_i\to m$.

Similarly the fermion eigenvalues are
\bea\label{eq:messengermassf}
m_{{5}_M} &= \frac{1}{2^{3/2}} \lp \lambda x + \sqrt{(\lambda x)^2 + 8 m^2} \rp,\nn\\ 
m_{{\overline{5}_M}} &= \frac{1}{2^{3/2}} \lp \lambda x - \sqrt{(\lambda x)^2 + 8 m^2} \rp.
\eea 

\textbf{$\bm{U(1)_D}$ gauge mediators.} The mass matrix of the $\phi$ fields is identical to that of the $5_M$ with the exception of the additional Fayet-Iliopoulos contribution. This doesn't affect the fermion mass matrix, so their mass eigenvalues is equal to Eq.\ (\ref{eq:messengermassf}).

Unfortunately, solving for the mass squared eigenvalues of the scalars requires finding the roots of a 4th order polynomial, which we cannot do analytically. In the limit $g\ll1$, those eigenvalues are also given by Eq.\ (\ref{eq:messengermassb}). We shall use this approximation to estimate the loop corrections to the scalar potential.

At leading order, we can simply shift the eigenvalues of the $SU(5)$ messengers by $\pm gD$. This is enough to see what happens in the limit of large $x$: some eigenvalues which previously converged to $m^2\to 0$ will instead converge to $m^2\to -gD$, that is, the model will have tachyons. At this point the potential becomes unstable at $\phi=\tilde\phi=0$ and the $U(1)$ messengers will get a VEV. This VEV will allow the scalar fields to cancel out the $D$ term in the limit $x\to\infty$. In other words, the true vacuum of the tree-level potential is the runaway solution $x\to\infty$. A slightly more accurate estimate of the mass eigenvalues is given by Eqs.\ (4.17)-(4.19) in Craig et al.

\textbf{Large $x$ approximation.} For large values of the pseudomodulus field $x$, most of the masses become either very large ($\sim x$) or very small ($\sim 1/x$). This allows one to derive simple approximations for the vacuum and thermal corrections (\ref{eq:CWpot},\ref{eq:thermalpot}). To simplify the analysis, we will only consider the contribution from the messengers. This is justified since they have a much larger number of degrees of freedom than the neutrinos. Furthermore, we will use the analytical formulas (\ref{eq:messengermassb}-\ref{eq:messengermassf}) for all the messenger masses, including the $\phi$.

Summing Eq.\ (\ref{eq:CWpot}) over all the mass eigenstates (\ref{eq:messengermassb}-\ref{eq:messengermassf}) and expanding to lowest order in $1/x$, one obtains
\be
V_{CW}(x\rightarrow\infty) \cong \frac{3\lambda^2F^2}{8\pi^2}\log\lp\frac{\lambda^2x^2}{2m^2}\rp,
\ee
where we neglected constant terms. The calculation of the thermal potential is slightly different, as large mass eigenstates are Boltzmann suppressed. One can therefore only consider the light eigenstates with the high-temperature approximation (\ref{eq:highT}). The leading-order thermal correction is then
\be
V_T(x\rightarrow\infty,T) \cong \frac{3m^4T^2}{\lambda^2 x^2}.
\ee

\end{appendix}

\bibliography{ref}
\bibliographystyle{utphys}
\end{document}